\documentclass[10pt,journal,compsoc]{IEEEtran}
\usepackage{booktabs} % For formal tables
\usepackage{amsmath}
\usepackage{subfigure}
\usepackage{tabularx}
\usepackage{graphicx}
\usepackage{algorithmicx, algorithm}
\usepackage[noend]{algpseudocode}

\usepackage{cuted}
\usepackage{capt-of}
\usepackage{color}
\usepackage[table]{xcolor}
\usepackage{booktabs}
\usepackage{lineno}

\newcommand*{\highlight}{\textcolor{black}}

\ifCLASSOPTIONcompsoc
  \usepackage[nocompress]{cite}
\else
  \usepackage{cite}
\fi

\ifCLASSINFOpdf
\else
\fi

\begin{document}
%\linenumbers
\title{A Variational Staggered Particle Framework for Incompressible Free-Surface Flows}

\author{Xiaowei He,
        Huamin Wang,
        Guoping Wang,
        Hongan Wang
        ~and~Enhua Wu% <-this % stops a space
\IEEEcompsocitemizethanks{
\IEEEcompsocthanksitem X. He is with the State Key Lab. of CS, Institute of Software, Chinese Academy of Sciences.\protect\\
E-mail: xiaowei@iscas.ac.cn
\IEEEcompsocthanksitem H. Wang is with the Ohio State University.\protect\\
E-mail: whmin@cse.ohio-state.edu
\IEEEcompsocthanksitem G. Wang is with the Peking University. \protect\\
E-mail: wgp@pku.edu.cn
\IEEEcompsocthanksitem H. Wang is with Institute of Software, Chinese Academy of Sciences. \protect\\
E-mail: hongan@iscas.ac.cn
\IEEEcompsocthanksitem E. Wu is with Institute of Software, Chinese Academy of Sciences and the University of Macau. \protect\\
E-mail: ehwu@umac.mo
}% <-this % stops an unwanted space
\thanks{Manuscript received X X, XXXX; revised X X, XXXX.}}
\IEEEtitleabstractindextext{%
\begin{abstract}
Smoothed particle hydrodynamics (SPH) has been extensively studied in computer graphics to animate fluids with versatile effects.
\highlight{However, SPH still suffers from two numerical difficulties: the particle deficiency problem, which will deteriorate the simulation accuracy, and the particle clumping problem, which usually leads to poor stability of particle simulations.
We propose to solve these two problems by developing an approximate projection method for incompressible free-surface flows under a variational staggered particle framework.
After particle discretization, we first categorize all fluid particles into four subsets.
Then according to the classification, we propose to solve the particle deficiency problem by analytically imposing free surface boundary conditions on both the Laplacian operator and the source term.
To address the particle clumping problem, we propose to extend the Taylor-series consistent pressure gradient model with kernel function correction and semi-analytical boundary conditions.
Compared to previous approximate projection method~\cite{He:2012:SMS}, our incompressibility solver is stable under both compressive and tensile stress states, no pressure clumping or iterative density correction (e.g., a density constrained pressure approach) is necessary to stabilize the solver anymore.}
Motivated by the Helmholtz free energy functional, we additionally introduce an iterative particle shifting algorithm to improve the accuracy.
It significantly reduces particle splashes near the free surface.
Therefore, high-fidelity simulations of the formation and fragmentation of liquid jets and sheets are obtained for both the two-jets and milk-crown examples.
  \end{abstract}

\begin{IEEEkeywords}
  particle deficiency, nonlocal, smoothed particle hydrodynamics, incompressibility, tensile instability.
\end{IEEEkeywords}}

\maketitle

\IEEEdisplaynontitleabstractindextext


% \fi
\IEEEpeerreviewmaketitle

%\IEEEraisesectionheading{\section{Introduction}\label{sec:introduction}}
%\IEEEPARstart{G}ranular media, after water, is the second-most-manipulated substance by man on Earth~\cite{De:1999:GMT}.

\section{Introduction}
\begin{figure}[t]
	\centering
	\includegraphics[width=1.0\linewidth]{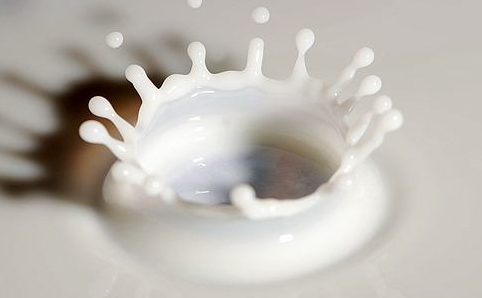}
	\vspace{-0.1in}
	\caption{\label{fig:photo} A photograph capturing the milk crown with viscous fingering structures resulted from the interaction among pressure, viscosity and surface tension.}
\end{figure}

Due to the meshless, Lagrangian nature, particle methods have been commonly used in computer graphics to animate incompressible free-surface flows.
In general, particle methods applied for free-surface flows can be categorized into two families: one is based on an equation of state (EOS), which either uses a non-iterative strategy~\cite{Becker:2007:WCS} or an iterative one~\cite{Solenthaler:2009:PIS,He:2012:LPS,Macklin:2013:PBF}, and the other is based on projection, which either tries to solve a constant density field~\cite{Ihmsen:2014:IIS} or a divergence-free velocity field~\cite{Bender:2015:DFS}.
Although those studies have shown a promising potential of particle methods in creating large-scale splashing fluids, the full exploitation of particle methods in creating subtle effects of fluids is still hampered by numerical problems involving both inaccuracy and instability.
One obvious example that is difficult for above mentioned methods to create is the \textit{viscous fingering} effect in the milk splash, as demonstrated in Figure~\ref{fig:photo}.

Two numerical problems that have long plagued particle methods in modeling incompressible free-surface flows are the particle deficiency and the particle clumping problems~\cite{Gotoh:2016:CAF}.
The particle deficiency problem is an issue where only particles inside the boundary contribute to the summation of particle interactions for particles near the free-surface boundary.
The missing particles usually will have a negative impact on simulation accuracy~\cite{Ikari:2015:CHO}.
Furthermore, since we only store a finite and often small number of particle neighbors in real implementation, the particle deficiency problem can also arise for interior particles when the particle distribution becomes irregular during the simulation~\cite{Colagrossi:2012:PPA}.
The particle clumping problem is a situation where particles may unnaturally cluster together resulted from a combined action of stress states and kernel functions~\cite{Liu:2010:SPH}.
For some historical reasons, the particle clumping problem arising from a tensile stress state is usually referred to as tensile instability~\cite{Swegle:1995:SPH,Monaghan:2000:SPH} while that arising from a compressive stress state referred to as pairing instability~\cite{Schuessler:1981:CSP,Dehnen:2012:ICS}.
Nevertheless, some researchers in engineering do not make an explicit distinction between tensile instability and pairing instability, e.g., Sugiura and Inutsuka~\cite{Sugiura:2016:EGS} refer to both instabilities as the tensile instability.
Since there is no naming unification for the particle clumping problem yet and the term `tensile' is sometimes misleading, we will use the term tensile instability specifically for the particle clumping problem arising from negative stress regimes and pairing instability for the particle clumping problem arising from positive stress regimes in the following discussion.
A sufficient condition for checking unstable growth of tensile instability in terms of the stress state and the second derivative of smoothing function was first proposed by Swegle et al.~\cite{Swegle:1995:SPH}.
However, Dehnen and Aly~\cite{Dehnen:2012:ICS} later disproved their statement by pointing out that the Wendland kernels does not suffer from the pairing instability despite having vanishing derivative at the origin.
To avoid the particle clumping problem, researchers in computer graphics have either clamped negative pressures to zero~\cite{Ihmsen:2014:IIS,Bender:2015:DFS} or added an artificial pressure~\cite{He:2014:RSS} to remove cohesive forces.
Unfortunately, the cohesive force imposed on boundary particles, which are the key to create the viscous fingering effect, will also be removed.
According to Belytschko and Xiao~\cite{Belytschko:2002:SAP}, perfect elimination of tensile instability appears to be unachievable as long as an Eulerian kernel is used with a purely Lagrangian description of motion.
Nevertheless, it is possible for us to minimize the influence of tensile instability by improving the accuracy of particle methods, e.g., by resolving the particle deficiency problem as mentioned earlier or selecting an appropriate kernel function.

Motivated by the variational framework~\cite{Batty:2007:FVF} and recent developments on nonlocal methods~\cite{He:2012:SMS,Du:2013:NVC,Ganzenmuller:2015:SMD}, we first reformulate incompressible free-surface flows as an energy minimization problem under a variational staggered particle framework.
Although the equivalence between solving a pressure Poisson equation and an energy minimization problem can be easily established for a uniform-grid based discretization, their equivalence is not quite obvious for a particle discretization.
\highlight{To get a stable and accurate simulation based on particles, both the discretized pressure Poisson equation and the pressure forces should be derived by meticulously addressing the particle deficiency and particle clumping problems.}
During the derivation, we have made the following contributions
\begin{itemize}
	\item{A reformulation of the particle discretized pressure Poisson equation derived from a variational staggered particle framework for incompressible free-surface flows.}
	\item{A new semi-analytic strategy to impose free-surface boundary conditions on both the Laplacian operator and the source term.}
	\item{An extended Taylor-series consistent pressure gradient model that is stable under both compressive and tensile stress states.}
	\item{An iterative particle shifting algorithm motivated by the Helmholtz free energy functional that not only helps regularize particle distributions, but also capture realistic surface tension effects. }
\end{itemize}

%\begin{figure}[t]
%\centering
%\includegraphics[width=1.0\linewidth]{images/photo.pdf}
%\vspace{-0.1in}
%  \caption{\label{fig:photo} A photograph capturing the milk splash with viscous fingering effects resulted from the interaction among pressure, viscosity and surface tension.}
%\end{figure}
\begin{figure}[t]
	\centering
	\includegraphics[width=1.0\linewidth]{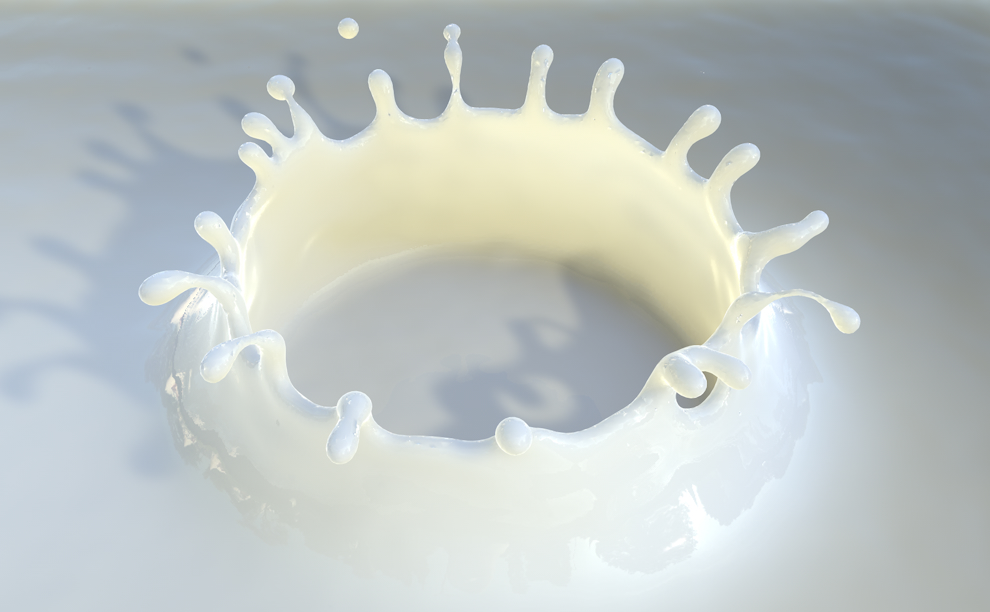}
	\vspace{-0.1in}
	\caption{\label{fig:milkcrown91} A milk crown generated by our method and realistic fingering structures can be noticed around the rim of the milk crown.}
\end{figure}

\section{Related Work}
Since SPH was first invented by Lucy~\cite{Lucy:1977:NAT} and Gingold and Monaghan~\cite{Gingold:1977:SPH}, various techniques have been proposed to solve the fluid incompressibility.

\textbf{EOS-based methods}. \highlight{Before SPH was first introduced into computer graphics by Desbrun and Gascuel~\cite{Desbrun:1996:SPA}, it had been applied in engineering to solve a wide range of dynamical problems (please see~\cite{Monaghan:2005:SPH} for a review).}
Researchers in computer graphics initially applied an equation of state (EOS) based on either a gas equation \cite{Muller:2003:PFS} or Tait's equation~\cite{Monaghan:1992:SPH} to model weakly compressible fluids.
To remove time step restrictions, Solenthaler and Pajarola~\cite{Solenthaler:2009:PIS} proposed a predictive-corrective incompressible SPH (PCISPH) to correct the density error iteratively.
Their method can handle time steps which are up to two orders of magnitude larger than previous non-iterative EOS-based methods.
He and colleagues \cite{He:2012:LPS} pointed out that the convergence rate of an iterative method is closely related to a particle's influence domain and proposed a local Poisson SPH (LPSPH) method to solve the incompressibility.
Bodin et al.~\cite{Bodin:2012:CF} enforced the incompressibility by solving a system of velocity constraints.
Recently, Macklin and M{\"u}ller \cite{Macklin:2013:PBF} presented an iterative density solver based on position based dynamics, which showed a better stability and performance than previous iterative methods.
For a thorough review, we refer to the work~\cite{Ihmsen:2014:SPH}.
In case readers are interested in recent progress of SPH in engineering, please refer to other review papers~\cite{Liu:2010:SPH,Violeau:2016:SPH}.

\textbf{Projection-based methods}.In solving the pressure Poisson equation, both the Laplacian operator and the source term should be discretized.
According to~\cite{Furstenau:2017:CNS}, the  most commonly used way to discretize the PPE in SPH is the finite difference scheme introduce by~\cite{Cummins:1999:SPM}, which is usually referred to as the approximate projection method.
Along this direction, Shao et al~\cite{Shao:2003:ISM} presented an ISPH method to simulate Newtonian and non-Newtonian flows with free surfaces.
He et al.~\cite{He:2012:SMS} proposed an approximate projection method based on staggered particles to solve the zero-energy mode problem, but leaves the particle deficiency problem untouched.
Nair and Gaurav~\cite{Nair:2014:IFS} presented a semi-analytical approach to impose a zero pressure boundary condition on free surfaces, achieving to an improved Laplacian operator for the pressure Poisson equation.
Yang et al.~\cite{Yang:2016:ESS} further improved the accuracy by considering the particle deficiency for the pressure force formulation.

\highlight{Other methods use the double summation scheme for discretization, which can be referred to as the exact projection method.
Hu and Adams~\cite{Hu:2009:CDA} proposed to correct intermediate density errors by adjusting the half-time-step velocity with exact projection for incompressible multi-phase SPH.
Ihmsen et al.~\cite{Ihmsen:2014:IIS} presented an implicit incompressible SPH (IISPH) to unilaterally enforce the incompressibility.
Compared to the approximate projection method, the exact projection method is typically regarded as less stable, e.g., it can suffer from oscillations and zero-energy mode if negative particle pressures are not clamped to zero~\cite{Cummins:1999:SPM}.
Later, Bender and Koschier~\cite{Bender:2015:DFS} improved the IISPH method by iteratively enforcing the divergence-free condition.
Band et al.~\cite{Band:2018:PBI} improved the solid wall boundary condition for IISPH.
Cornelis et al~\cite{Cornelis:2019:OST} presented an analysis of two source terms and proposed to incorporate velocity divergence and particle shift to reduce artificial viscosity.}

Alternatively, the pressure Poisson equation can be solved on a regular grid~\cite{Losasso:2008:TWC,Raveendran:2011:HSP} with a similar idea to FLIP~\cite{Zhu:2005:ASF}, but at a cost of losing the purely Lagrangian nature of SPH.
\highlight{For more details about recent developments on projection-based particle methods in engineering, please refer to the work~\cite{Gotoh:2016:CAF}.}

\textbf{Inherent numerical problems}.\highlight{The \nobreak{development} of projection-based SPH methods in computer graphics is much slower than the EOS-based methods.
The reason is that the pressure Poisson equation is sensitive to the underlying particle distribution and suffers from numerical problems involving particle deficiency and tensile instability.}
For a truly incompressible fluid, the negative pressures should not be simply removed. Therefore, a robust and accurate fluid solver for incompressible free-surface flows is required.
\highlight{To our best knowledge, only little work has been done in computer graphics on how to solve the above mentioned numerical problems.}
Schechter and Bridson~\cite{Schechter:2012:GSA} added ghost air particles to help resolve the particle deficiency problem, but it requires a significant extra computational cost.
Macklin and M{\"u}ller~\cite{Macklin:2013:PBF} and He et al~\cite{He:2014:RSS} added an artificial pressure to alleviate the tensile instability, which is equivalent to adding an artificial surface tension or removing negative pressures, respectively.
However, none of these techniques work well for the projection methods.
In engineering, the numerical problems have also been extensively studied~\cite{Gotoh:2016:CAF}.
However, as pointed out in their work, both the stability and accuracy of particle methods have not yet fully addressed.

\section{A Variational Staggered Particle Framework}
In the context of projection-based methods, the incompressibility of a free-surface flow is enforced by solving the following pressure Poisson equation
\begin{linenomath*}
	\begin{equation}
		\begin{array}{l}
			\begin{aligned}
				\nabla  \cdot \left( {\frac{{\Delta t}}{\rho }\nabla {p}} \right) = \nabla  \cdot {{\bf{v}}^*}, {\kern 15pt} &inside{\kern 3pt}  \Omega, \\
				p = 0, {\kern 16pt} &on{\kern 3pt}  \partial \Omega,
			\end{aligned}
		\end{array}
		\label{eq:poisson}
	\end{equation}
\end{linenomath*}
where $\Omega$ is the fluid region with free surface boundary $\partial \Omega$, $p$ is the pressure, $\Delta t$ is the time step, $\rho$ is the density and ${\bf{v}}^*$ represents the intermediate velocity which has considered all forces except the pressure force.

\begin{figure}[t]
	\centering
	\subfigure[Local]{\includegraphics[width=0.48\linewidth]{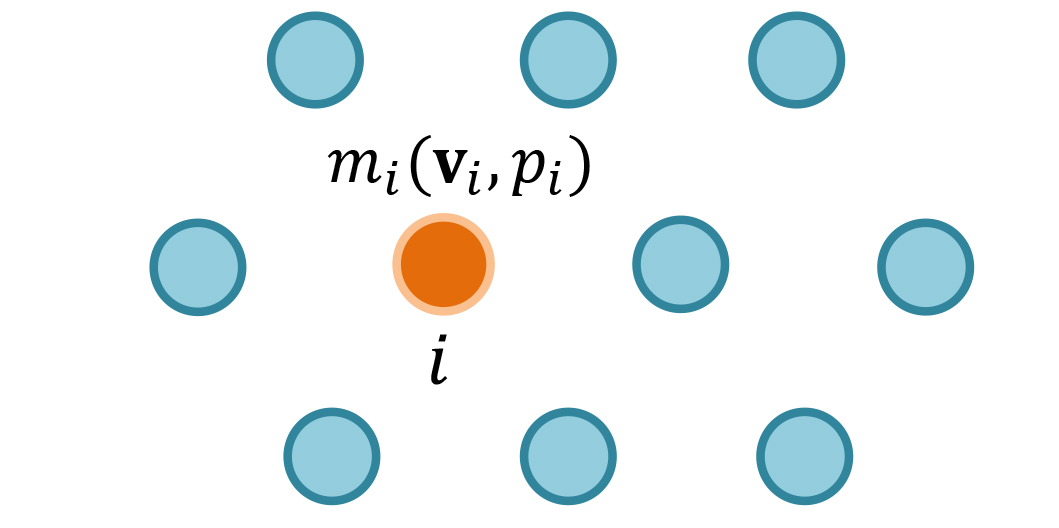}}
	\subfigure[Nonlocal]{\includegraphics[width=0.48\linewidth]{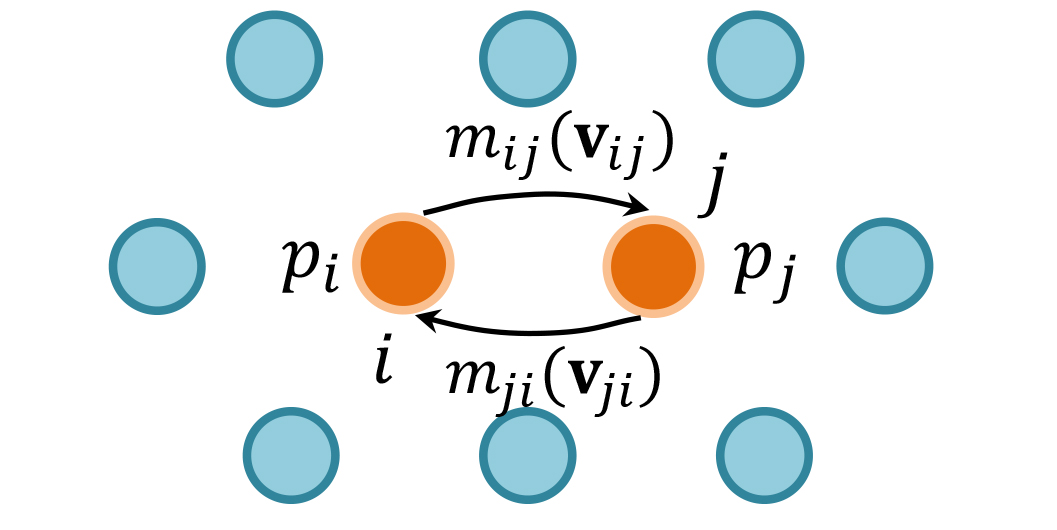}}
	\vspace{-0.1in}
	\caption{\label{fig:discretization} Local vs nonlocal variational frameworks. (a) Under a local variational framework, all physical quantities are carried on the same particles; (b) Under our variational staggered particle framework, particle masses and velocities are decoupled from original particles and defined as nonlocal variables.}
\end{figure}
\subsection{Motivation}
According to~\cite{Batty:2007:FVF}, the pressure Poisson equation can be reformulated as an energy minimization problem
\begin{linenomath*}
	\begin{equation}
		\mathop {\min }\limits_{p} \int_\Omega  {\frac{1}{2}\rho {{\left\| {\mathbf{v}^* - \frac{{\Delta t}}{\rho }\nabla p} \right\|}^2} dV}, {\kern 13pt}  p = 0 {\kern 3pt} on{\kern 3pt} \partial \Omega .
		\label{eq:Ek}
	\end{equation}
\end{linenomath*}
Intuitively speaking, enforcing a fully incompressible fluid is equivalent to maximally dissipating the kinetic energy by using the pressure force.
The question is how can we formulate the energy minimization problem in SPH?

Figure~\ref{fig:discretization}a demonstrates the standard discretization for a fluid region in SPH, where each particle $i$ carries a set of local physical quantities including the particle mass $m_i$, velocity $\mathbf{v}_i$ and pressure $p_i$, etc.
We call the quantities to be \emph{local} because their values only depend on the position of particle $i$.
In this sense, the pressure gradient $\nabla p$ estimated in traditional SPH is also a local quantity.
Inserting all local quantities into Equation~\ref{eq:Ek}, a local form of discretization for the energy minimization problem can be derived as
\begin{linenomath*}
	\begin{equation}
		\mathop {\min }\limits_{{p}} \sum\limits_{i,j} {\frac{1}{2}m_i {{\left\| {{\mathbf{v}_i^*} - \frac{{\Delta t}}{\rho_0 }\sum \nolimits_j {\frac{{{m_j}}}{{{\rho _j}}}\left( {\frac{{{p_i} + {p_j}}}{2}} \right){\nabla _i}{W}} } \right\|}^2}},
		\label{eq:localEk}
	\end{equation}
\end{linenomath*}
where $p_i$ is zero for air boundary particles, $W$ is the kernel function, $\rho_0$ is the reference density, $j$ represents all neighbors of particle $i$.
Taking the derivative of Equation~\ref{eq:localEk} with respect to $p_i$, we obtain a discretized pressure Poisson equation that corresponds to an exact projection where the second-ring neighbors of particle $i$'s neighbors should also be taken into account for projection.
As pointed in~\cite{Cummins:1999:SPM}, the major problem with the exact SPH projection method is that it suffers from the spurious zero-energy mode problem, which means the pressure field could be oscillating unnaturally.
The underlying reason is that both the pressure and velocity are local variables and defined at the same location.
He and his colleagues~\cite{He:2012:SMS} propose to address the zero-energy mode problem by introducing staggered particles.
The core idea of their method is to decouple the definition of the pressure and other physical quantities and define their values at different locations.
This inspires us to introduce nonlocal quantities and reformulate the energy minimization problem in Equation~\ref{eq:Ek} to be a staggered form in the following section.

\subsection{Our Discretization}
Before deriving the variational staggered particle framework, we introduce two types of functions first:
point functions that refer to functions defined on single particles and two-point functions that refer to functions defined for pairs of particles, e.g., we define $\psi \left( {{\mathbf{x}_i}} \right)$ as local scalars and $\psi \left( {{\mathbf{x}_i}}, {\mathbf{x}_j} \right)$ as nonlocal scalars.
We refer to the work of Du et al~\cite{Du:2013:NVC} for more details on the discussion of nonlocal operators.
Figure~\ref{fig:discretization}b demonstrates a staggered discretization strategy that both the mass and velocity are defined as two-point functions
\begin{linenomath*}
	\begin{equation}
		{{m}_{ij}} = \underline{m}\left( {{\mathbf{x}_i},{\mathbf{x}_j}} \right),{\kern 10pt}{{\mathbf{v}}_{ij}} = \underline{\mathbf{v}}\left( {{\mathbf{x}_i},{\mathbf{x}_j}} \right),
		\label{eq:newmv}
	\end{equation}
\end{linenomath*}
where the underline is used to distinguish two-point functions from point functions.
Besides, we assume the pressure force imposed on $m_{ij}$ is only related to particle $i$ and $j$, as was done in~\cite{He:2012:SMS}, the pressure gradient imposed on $m_{ij}$ is expressed as follows
\begin{linenomath*}
	\begin{equation}
		{\nabla _{ij}}p = \frac{{{p_j} - {p_i}}}{{{r_{ij}}}}{\mathbf{n}_{ij}},
		\label{eq:grad}
	\end{equation}
\end{linenomath*}
in which $r_{ij}=\left\| \mathbf{x}_j - \mathbf{x}_i \right\|$ and $\mathbf{n}_{ij}=\left( \mathbf{x}_j - \mathbf{x}_i \right)/r_{ij}$.
Inserting both Equation~\ref{eq:newmv} and~\ref{eq:grad} into Equation~\ref{eq:Ek}, we get the following staggered formulation of the energy minimization problem
\begin{linenomath*}
	\begin{equation}
		\mathop {\min }\limits_{{p}} \sum\limits_{i,j} {\frac{1}{2}{m_{ij}}{{\left\| {{{\bf{v}}_{ij}^*} - \frac{{\Delta t}}{\rho _{0}}\left( {\frac{{{p_j} - {p_i}}}{{{r_{ij}}}}} \right){\mathbf{n}_{ij}}} \right\|}^2}}.
		\label{eq:nonlocalEk}
	\end{equation}
\end{linenomath*}
The major difference between Equation~\ref{eq:nonlocalEk} and~\ref{eq:localEk} is that the pressure and velocity field are decoupled in equation~\ref{eq:nonlocalEk}, which is the key to solve the zero-energy mode problem.

Now we discuss how to calculate the nonlocal mass $m_{ij}$ and velocity $\mathbf{v}_{ij}$.
Assume the particle mass and velocity are initially stored as local variables $m_i$ and $\mathbf{v}_i$, we introduce the following two-point operators to map local variables to nonlocal ones
\begin{linenomath*}
	\begin{equation}
		{m_{ij}} = \frac{{{\omega _{ij}}}}{\alpha_i} m_i, {\kern 10pt} {\mathbf{v}_{ij}} = \frac{\mathbf{v}_i+\mathbf{v}_j}{2} ,
		\label{eq:twopoint}
	\end{equation}
\end{linenomath*}
where $\omega _{ij}$ is a weighting function required to be symmetric, i.e., $\underline{\omega}(\mathbf{x}_i, \mathbf{x}_j)=\underline{\omega}(\mathbf{x}_j, \mathbf{x}_i)$.
By temporarily ignoring the particle deficiency problem, we are able to define the value of $\alpha_i$ to be $\alpha_i = {\sum\nolimits_j {{\omega _{ij}}} }$, where the total mass and momentum can be verified to be conservative after the mapping.
Note that the masses $m_{ij}$ and $m_{ji}$ may not be equal due to irregular particle distributions.
After inserting Equation~\ref{eq:twopoint} into Equation~\ref{eq:nonlocalEk}, we get the following discretized pressure Poisson equation by taking the derivative with respect to $p_i$
\begin{linenomath*}
	\begin{equation}
		\begin{array}{l}
			\begin{aligned}
				\sum\limits_j & {\frac{{1}}{\rho _{0}} \left( {\frac{1}{{{\alpha _i}}} + \frac{1}{{{\alpha _j}}}} \right) { \frac{{{p_i} - {p_j}}}{{r_{ij}^2}}{\omega _{ij}}} }  \\
				&{\kern 40pt} = \frac{1}{\Delta t} \sum\limits_j \left( {\frac{1}{{{\alpha _i}}} + \frac{1}{{{\alpha _j}}}} \right) {\left( {\frac{\mathbf{v}_i^* + \mathbf{v}_j^*}{2}} \right)} \cdot \frac{{{\mathbf{n}_{ij}}}}{{{r_{ij}}}}{\omega _{ij}},
			\end{aligned}
		\end{array}
		\label{eq:p2}
	\end{equation}
\end{linenomath*}
in which we have assumed all particles carry equal masses.
The left-hand side of Equation~\ref{eq:p2} now represents an approximate Laplacian operator while the right-hand side represents a source term.
More details on how to correct the value of $\alpha_i$ as well as the discretized pressure Poisson equation to account for the particle deficiency problem will be later discussed in section 4.

\begin{figure}[t]
	\centering
	\subfigure[]{\includegraphics[width=0.49\linewidth]{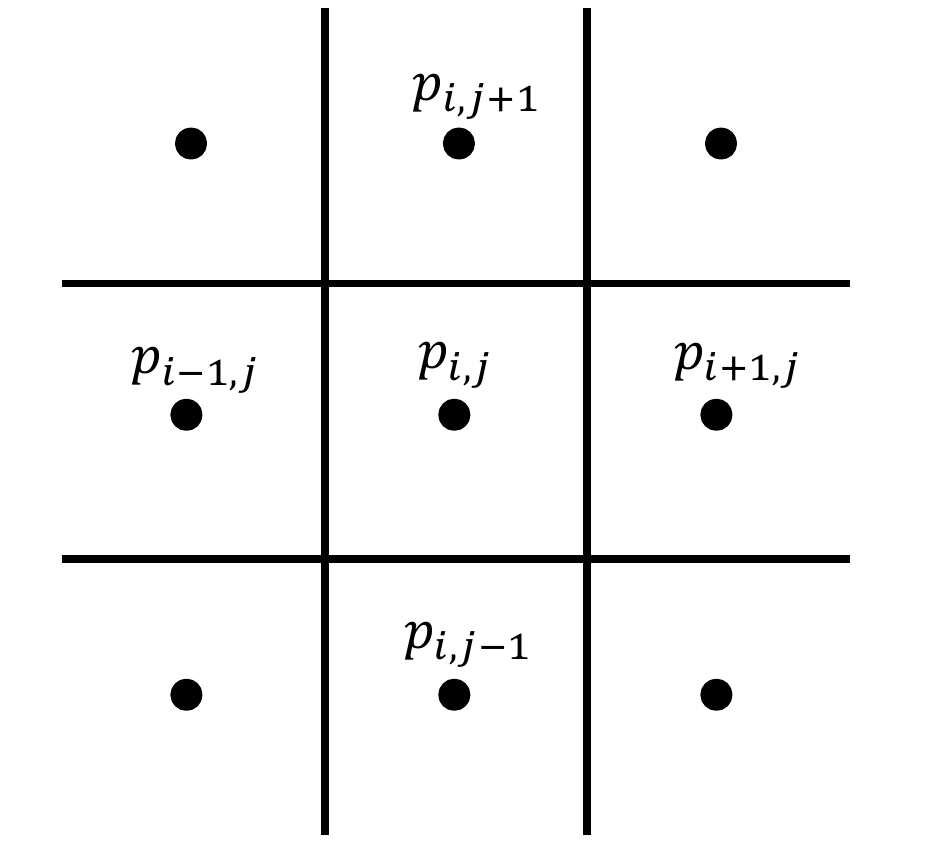}}
	\subfigure[]{\includegraphics[width=0.49\linewidth]{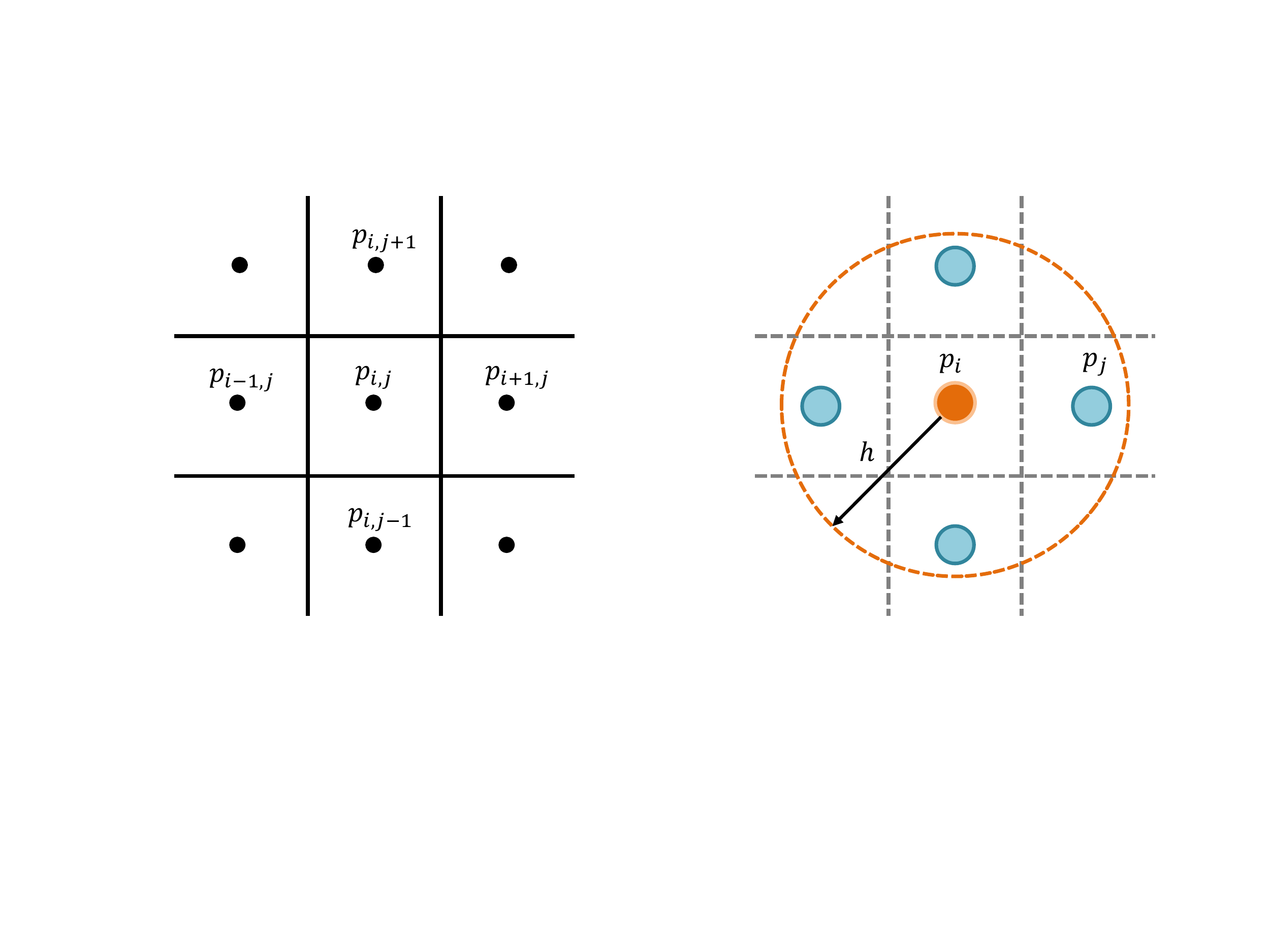}}
	\caption{\label{fig:laplacian}(a) The two-dimensional MAC grid, (b) An uniform particle distribution with four neighbors. The particle-based Laplacian operator proposed by [Cummins and Rudman 1999] is identical to the grid-based Laplacian operator only when we pick an exponential function, e.g., $W = e^{- {\rm ln} r}$, to be the kernel function.}
\end{figure}
\textbf{Comparison to [Cummins and Rudman 1999]}. The Laplacian operator proposed by Cummins and Rudman~\cite{Cummins:1999:SPM} is defined as
\begin{linenomath*}
	\begin{equation}
		\sum\limits_j {\frac{{{m_j}}}{{{\rho _j}}}\left( {\frac{4}{{{\rho _i} + {\rho _j}}}} \right)\frac{{{p_i} - {p_j}}}{{r_{ij}^2}}\left( {{\mathbf{x}_i} - {\mathbf{x}_j}} \right)  \cdot {\nabla _i}{W_{ij}}}.
		\label{eq:cum}
	\end{equation}
\end{linenomath*}
In case the fluid is ideally incompressible, we can verify that Equation~\ref{eq:cum} is equal to the left hand side of Equation~\ref{eq:p2} only if we impose the following two constraints
\begin{linenomath*}
	\begin{equation}
		W_{ij} = {\omega _{ij}} =  - {r_{ij}} \frac{{\partial {W_{ij}}}}{{\partial {r_{ij}}}}.
		\label{eq:w_relationship}
	\end{equation}
\end{linenomath*}
\begin{linenomath*}
	\begin{equation}
		{\rho _i} =  {\rho _0} = \sum\limits_j {{m_j}{W_{ij}}}.
	\end{equation}
\end{linenomath*}
To satisfy the condition in Equation~\ref{eq:w_relationship}, the kernel function $W$ should be in the form of an exponential function, i.e., $W_{ij} = e^{- {\rm ln} r_{ij}}$.
Unfortunately, an exponential function is usually not a good kernel function for the projection-based methods, because the value of $({p_i} - {p_j})/r_{ij}^2$ can become too large if a neighbor $j$ is close to particle $i$, and thus may cause instabilities during the simulation.
Therefore, in traditional SPH methods, it is common to use kernel functions whose gradient is zero at the original point where the values of $W_{ij}$ and $-\partial W_{ij} /\partial r_{ij}$ are usually different.
Therefore, we need to meticulously calculate the pressure force to avoid tensile instabilities under compressive stress states~\cite{Tsuruta:2013:SND}, especially when the Laplacian and gradient operator are discretized in different ways~\cite{Ma:2016:RAS}.
Among all discretization strategies, it is really hard to tell which pair of the Laplacian and gradient operators is the best to solve the pressure Poisson equation.
\begin{figure}[t]
	\centering
	\subfigure[${\mathcal P}^b$]{\includegraphics[width=0.49\linewidth]{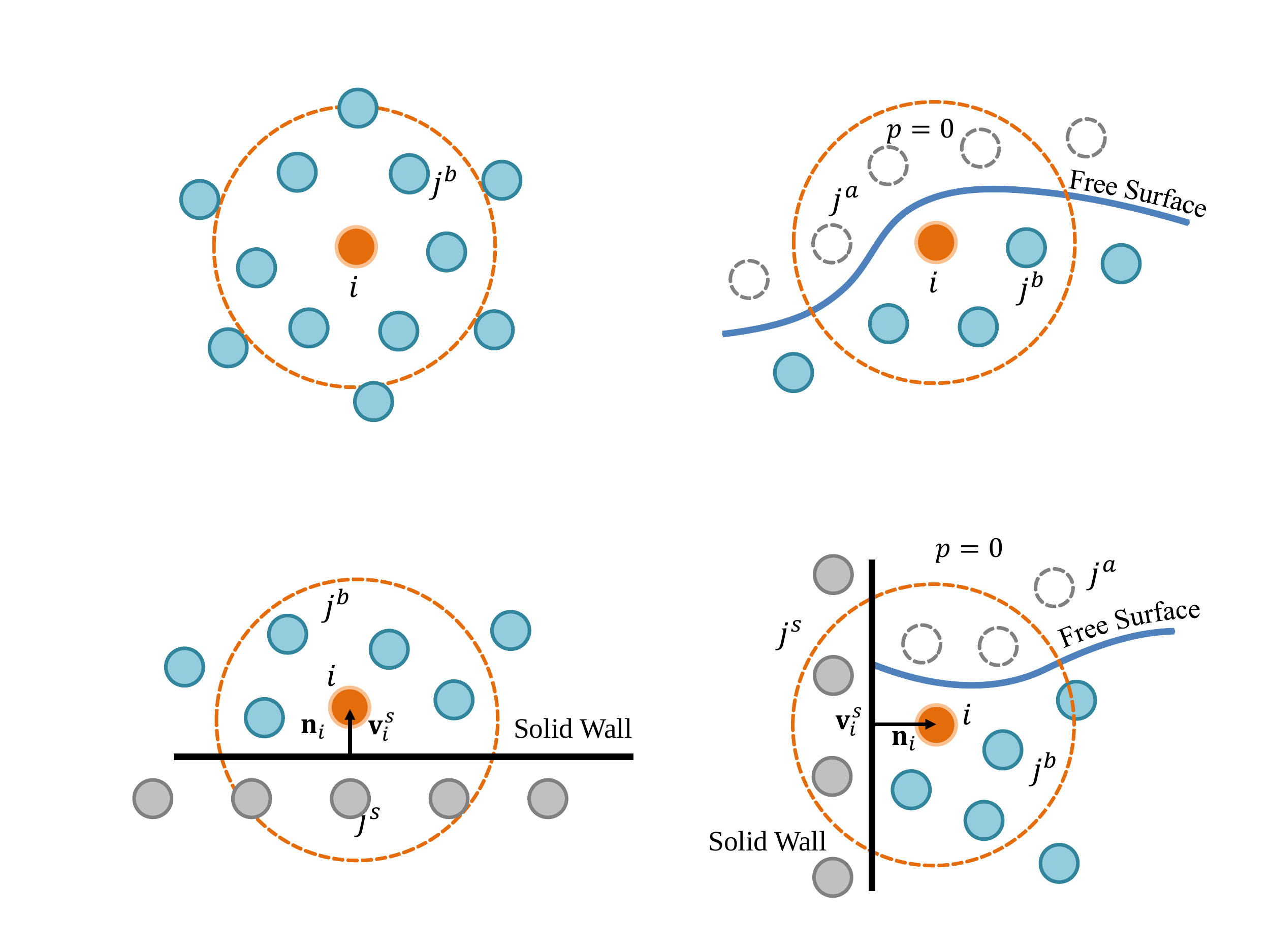}}
	\subfigure[${\mathcal P}^a$]{\includegraphics[width=0.49\linewidth]{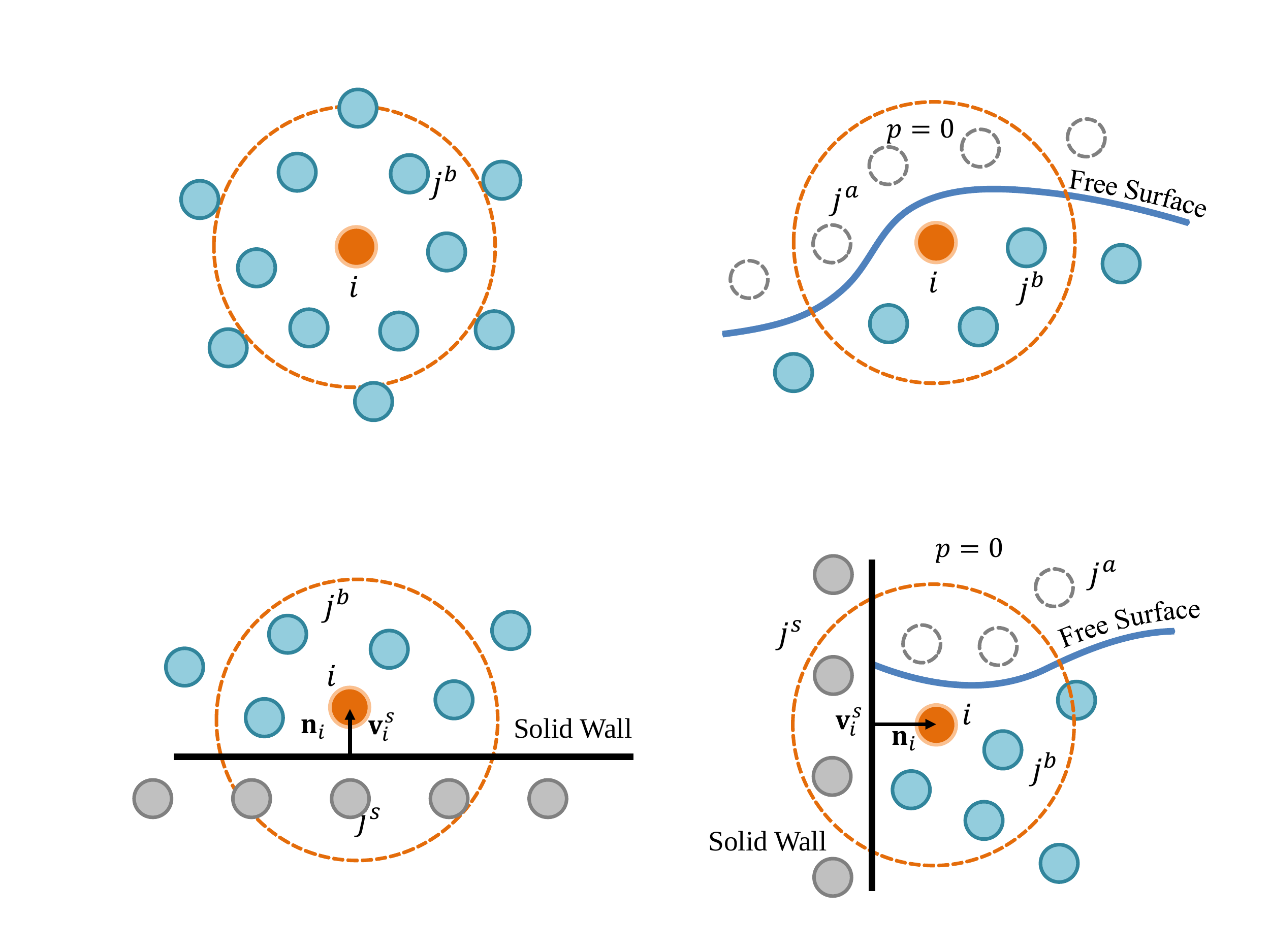}}
	\subfigure[${\mathcal P}^s$]{\includegraphics[width=0.49\linewidth]{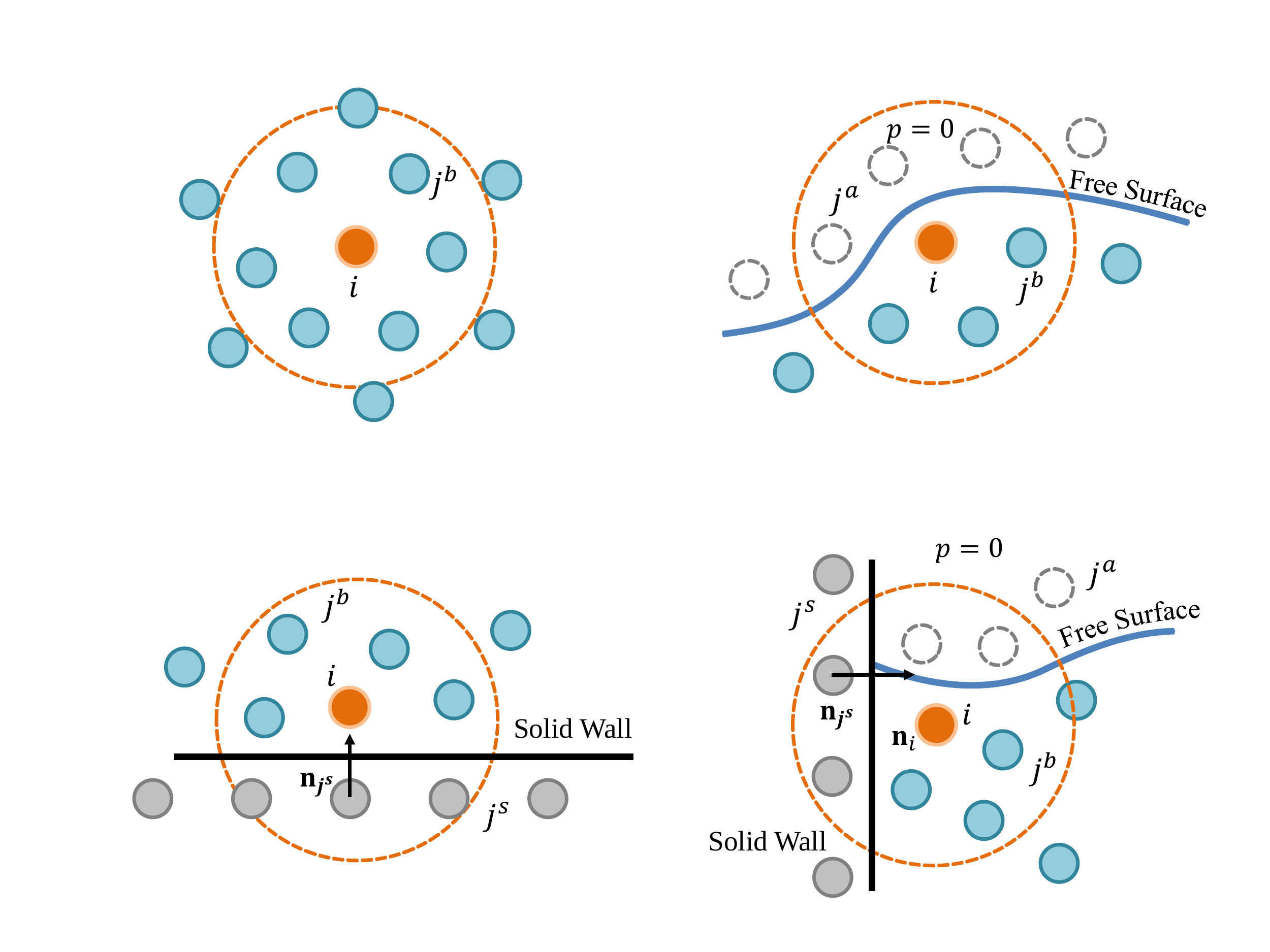}}
	\subfigure[${\mathcal P}^{a \wedge s}$]{\includegraphics[width=0.49\linewidth]{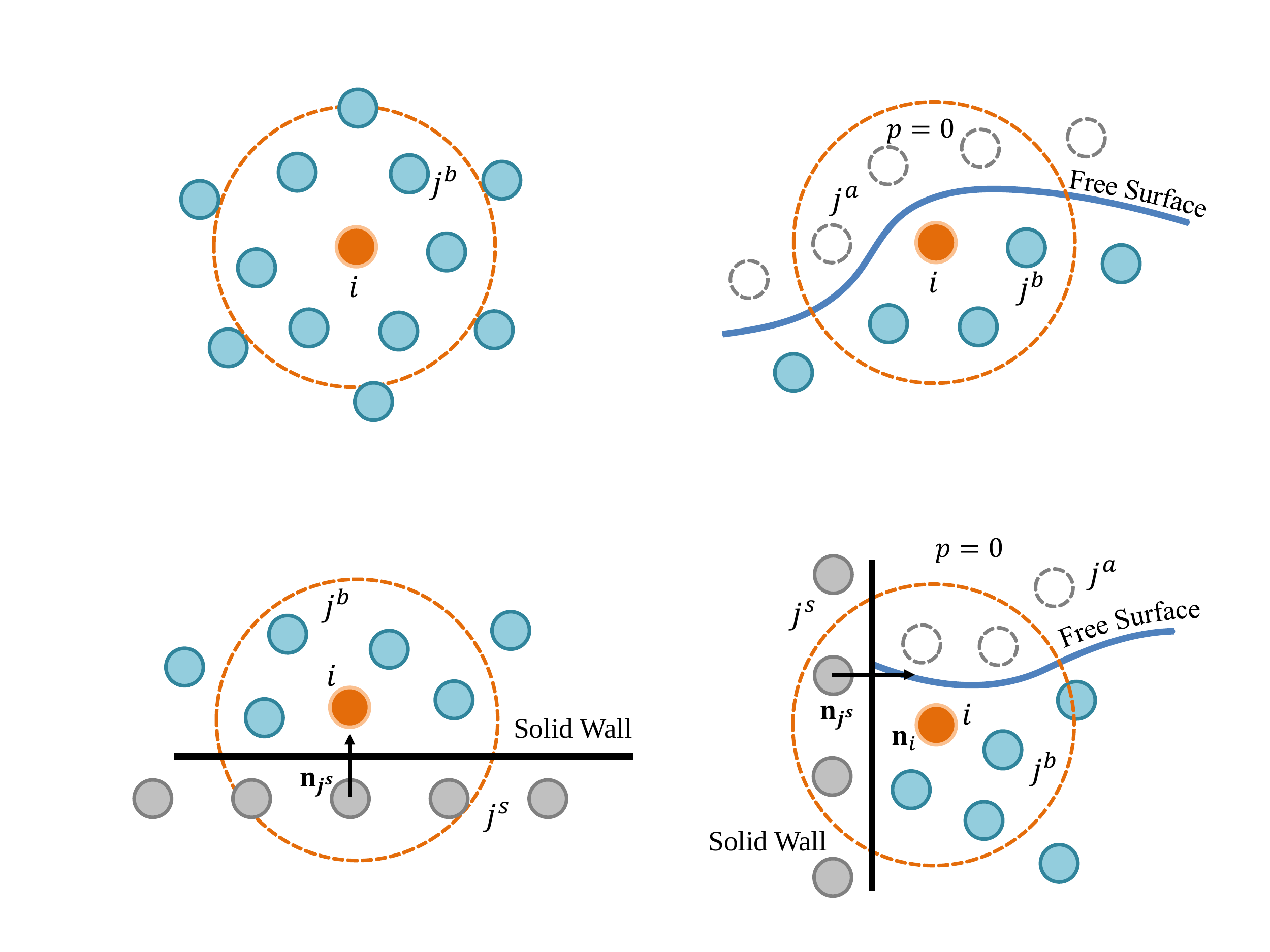}}
	\caption{\label{fig:fourcase} Illustration for our principle in categorizing all fluid particles into four subsets based on the intersection testing between a particle's support domain and fluid boundaries.}
\end{figure}
In fact, we can actually notice from Figure~\ref{fig:laplacian} that the particle-based Laplacian operator defined in Equation~\ref{eq:cum} does not even converge to an grid-based Laplacian operator if we choose an arbitrary kernel function.
Nevertheless, our discretization formulation does not suffer the above mentioned problems as both the Laplacian and gradient operators are derived uniformly under the same variational staggered particle framework.
Later, we will show how to resolve the particle clumping problem in section 5 by \highlight{selecting the right kernel function and introducing a correction for the kernel function}, therefore no adhoc tricks, e.g., the dynamic stabilization~\cite{Tsuruta:2013:SND}, are required to stabilize the PPE solver any more.
\begin{figure*}[t]
	\centering
	\subfigure[$\kappa = 0$]{\includegraphics[width=0.325\linewidth]{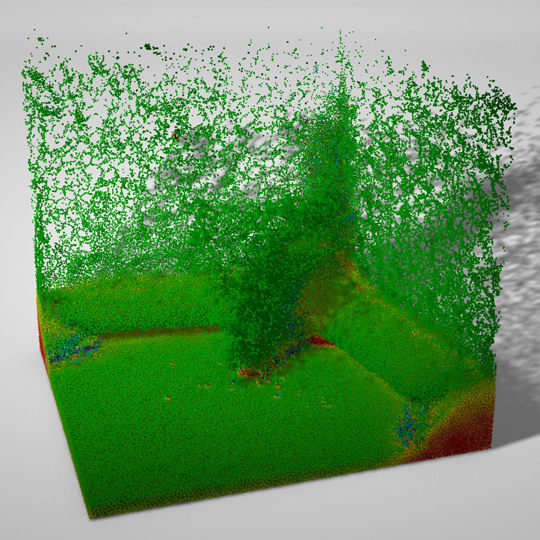}}
	\subfigure[$\kappa = 0.01$]{\includegraphics[width=0.325\linewidth]{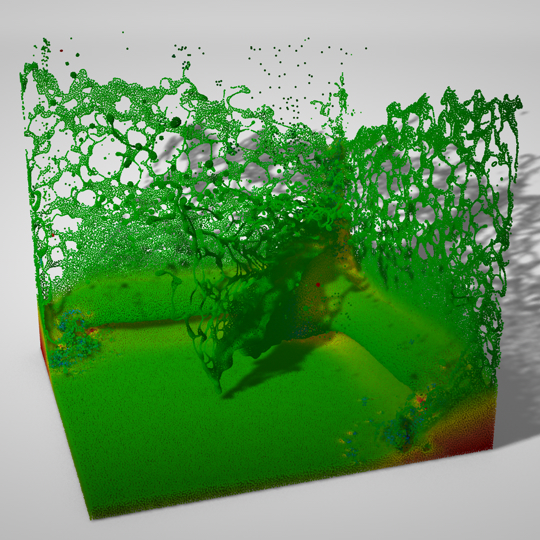}}
	\subfigure[$\kappa = 0.1$]{\includegraphics[width=0.325\linewidth]{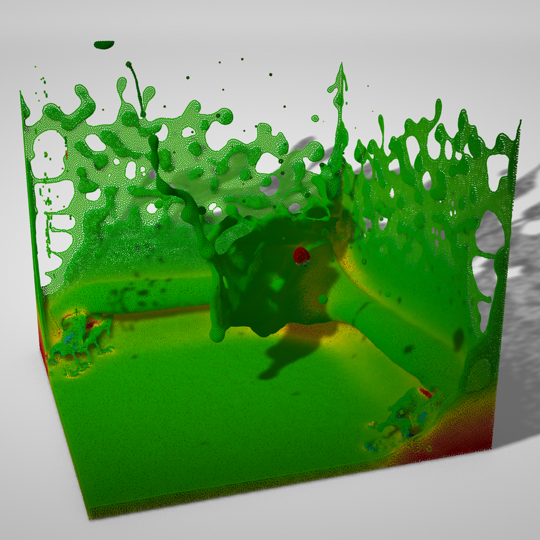}}
	\caption{\label{fig:dambreak} A three-dimensional dambreak case. This example shows the flow patterns of using different squared gradient energy coefficients. }
\end{figure*}

\section{Solving the Particle Deficiency Problem}
In this section, we will discuss how to address the particle deficiency problem for each part of Equation~\ref{eq:p2}.
In our implementation, ghost solid particles are uniformly seeded near the solid wall within a distance that equals to the smoothing length $h$ at the beginning of simulation.
Since dynamic creation of ghost air particles is time consuming, we propose a semi-implicit method, motivated by~\cite{He:2014:RSS,Nair:2014:IFS}, to virtually account for the contribution from ghost air particles.
To facilitate the following discussion, let us first assume that ghost air particles still exist, and denote ghost air neighbors as $j^a \in \mathcal{N}_i^a$, ghost solid neighbors as $j^s \in \mathcal{N}_i^s$ and fluid neighbors as $j^b \in \mathcal{N}_i^b$, thus we have $\mathcal{N}_i=\mathcal{N}_i^a \cup \mathcal{N}_i^b \cup \mathcal{N}_i^s$.

\subsection{Correcting the Laplacian operator}
Following the derivation of Bridson~\cite{Bridson:2015:FSC}, if a particle neighbor $j^a$ is a ghost air particle, we can set $p_{j^a}$ in Equation~\ref{eq:p2} simply to be zero.
Otherwise, if a particle neighbor $j^s$ is a ghost solid particle, \highlight{the velocity change $\Delta {\bf{v}}_{i{j^s}}^*$ of the staggered particle $m_{ij^s}$ caused by pressure force can be expressed as
\begin{linenomath*}
	\begin{equation}
		\Delta {\bf{v}}_{i{j^s}}^* = - \frac{{\Delta t}}{{{\rho _0}}}{\nabla _{i{j^s}}}p,
		\label{eq:update}
	\end{equation}
\end{linenomath*}
where ${\nabla _{i{j^s}}}p$ is the pressure gradient defined in Equation~\ref{eq:grad}.
However, to flexibly control the slipperiness of the solid wall boundary, it can be noticed that the velocity change $\Delta {\bf{v}}_{i{j^s}}^*$ can also be defined as the following constraint}~\cite{He:2012:SMS}
\begin{linenomath*}
	\begin{equation}
		\begin{array}{l}
			\begin{aligned}
				\Delta {\bf{v}}_{i{j^s}}^* = &{c^n}{\rm proj}_{{{\bf{n}}_{{j^s}}}}\left( {{{\bf{v}}_{{j^s}}} - {{\bf{v}}_i^*}} \right)  \\
				&+ {c^t}\left[ {\left( {{{\bf{v}}_{{j^s}}} - {{\bf{v}}_i^*}} \right) - {\rm proj}_{{{\bf{n}}_{{j^s}}}}\left( {{{\bf{v}}_{{j^s}}} - {{\bf{v}}_i^*}} \right)} \right],
			\end{aligned}
		\end{array}
		\label{eq:wall}
	\end{equation}
\end{linenomath*}
where $\mathbf{v}_{j^{s}}$ represents the velocity of the ghost solid neighbor $j^s$, ${{\bf{n}}_{{j^s}}}$ is a normal vector calculated from the signed distance field of the solid wall boundary, ${\rm proj}_{{{\bf{n}}_{{j^s}}}}( {{{\bf{v}}_{{j^s}}} - {{\bf{v}}_i^*}})$ represents the projection of ${{{\bf{v}}_{{j^s}}} - {{\bf{v}}_i^*}}$ on ${{\bf{n}}_{{j^s}}}$, $c^n$ and $c^t$ are two independent constants in the range of $[0, 1]$ to control the solid wall boundary condition.
Intuitively speaking, $c^t$ controls the sliding speed between particle $i$ and $j^s$.
Therefore, the value of $c^t$ can be set to 1 for a no-slip boundary condition and 0 for a free-slip boundary condition.
$c^n$ controls the normal speed between particle $i$ and $j^s$.
To prevent particle $i$ from interpenetrating into the solid wall, we will always set $c^n$ to 1 when particle $i$ approaches particle $j^s$ (i.e., $( {{{\bf{v}}_{{j^s}}} - {{\bf{v}}_i^*}}) \cdot {\bf{n}}_{j^s} > 0$).
Otherwise, we can control the strength of the fluid stickiness to the wall by adjusting the value of $c^n$.

After inserting all boundary conditions (please refer to Appendix for the whole derivation), the Laplacian operator in Equation~\ref{eq:p2} can be reformulated as follows
\begin{linenomath*}
	\begin{equation}
		{{\mathcal L}_i} = \frac{{{\hat A}_i}}{{{\rho _0}}}{p_i} - \frac{1}{{{\rho _0}}}\sum\limits_{{j^b}} {\left( {\frac{1}{{{{\hat \alpha }_i}}} + \frac{1}{{{{\hat \alpha }_j}}}} \right)\frac{{{\omega _{ij}}}}{{r_{ij}^2}}} {p_j},
		\label{eq:L2}
	\end{equation}
\end{linenomath*}
where
\begin{linenomath*}
	\begin{equation}
		\begin{array}{l}
			\begin{aligned}
				{{\hat A}_i} = \sum\limits_{{j^a \cup j^b}} {\left( {\frac{1}{{{{\hat \alpha }_i}}} + \frac{1}{{{{\hat \alpha }_j}}}} \right)\frac{{{\omega _{ij}}}}{{r_{ij}^2}}}
			\end{aligned}
		\end{array}
		\label{eq:para}
	\end{equation}
\end{linenomath*}
and ${\hat{\alpha} _i} = \sum\nolimits_{j} {{\omega _{ij}}}$ represents the total weight of a fluid particle with full neighbors.
The question is how can we calculate ${\hat \alpha}_i$ and ${\hat A}_i$ without knowing the locations of all ghost air particles in advance?

To avoid creating all ghost air particles $j^a$, we precompute two thresholds ${\alpha _0}$ and ${A_0}$ for an interior prototype particle with full fluid neighbors at the beginning of simulation as follows
\begin{linenomath*}
	\begin{equation}
		\begin{array}{l}
			\begin{aligned}
				{{\alpha} _0} = \sum\limits_{j} {{\omega _{ij}}},
				{\kern 10pt} {A_0} = \sum\limits_{{j}} {\left( {\frac{1}{{{{\alpha }_i}}} + \frac{1}{{{{\alpha }_j}}}} \right)\frac{{{\omega _{ij}}}}{{r_{ij}^2}}}
			\end{aligned}
		\end{array}
		\label{eq:A0}
	\end{equation}
\end{linenomath*}
In calculating $\hat{\alpha} _i$, we first compute the total weight by only considering contributions from neighbor particles $j^b$ and $j^s$, i.e., $\alpha^{b \wedge s}_i  = \sum\nolimits_{{j^b} \cup {j^s}} {{\omega _{ij}}}$.
Then, we calculate $\hat{\alpha} _i$ as follows
\begin{linenomath*}
	\begin{equation}
		\begin{array}{l}
			\begin{aligned}
				{\hat{\alpha} _{{i}}} = \max \left( {{{\alpha} _0}, \alpha^{b \wedge s}_i} \right)
			\end{aligned}
		\end{array}
		\label{eq:clamp}
	\end{equation}
\end{linenomath*}
to compensate for missing ghost air particles.
Equation~\ref{eq:clamp} also indicates that the value of $\alpha^{b \wedge s}_i$ could be occasionally larger than ${{\alpha} _0}$ during the simulation.
In that case, ${\hat{\alpha} _{{i}}}$ just equals to $\alpha^{b \wedge s}_i$.
Similarly, in calculating ${\hat A}_i$, we first calculate two terms from neighbors $j^b$ and $j^s$, which are expressed as
\begin{linenomath*}
	\begin{equation}
		A_i^b = \sum\limits_{{j^b}} {\left( {\frac{1}{{{{\hat \alpha }_i}}} + \frac{1}{{{{\hat \alpha }_j}}}} \right)\frac{{{\omega _{ij}}}}{{r_{ij}^2}}} , {\kern 10pt}A_i^s = \sum\limits_{{j^s}} {\left( {\frac{1}{{{{\hat \alpha }_i}}} + \frac{1}{{{{\hat \alpha }_j}}}} \right)\frac{{{\omega _{ij}}}}{{r_{ij}^2}}}.
		\label{eq:bs}
	\end{equation}
\end{linenomath*}
Before estimating the contribution from ghost air particles, we categorize all fluid particles into four subsets, as illustrated in Figure~\ref{fig:fourcase}:
\begin{enumerate}
	\item {$\mathcal{P}^{b}$: all interior particles whose support domain does not intersect with any of the boundaries, i.e., $\mathcal{N}_i^s = \emptyset$ and $A_i^b \ge A_0$.}
	\item {$\mathcal{P}^a$: all boundary particles whose support domain are only truncated by the free surface boundary, i.e., $\mathcal{N}_i^s=\emptyset$ and $A_i^b < A_0$; }
	\item {$\mathcal{P}^s$: all boundary particles whose support domain are only truncated by the solid wall boundary, i.e., $\mathcal{N}_i^s \ne \emptyset$ and $A_i^b + A_i^s \ge A_0$;}
	\item {$\mathcal{P}^{a \wedge s}$: all boundary particles whose support domain are both truncated by the free surface and solid wall boundaries, i.e., $\mathcal{N}_i^s \ne \emptyset$ and $A_i^b + A_i^s < A_0$;}
\end{enumerate}
By invoking the definition of $\hat A_i$ in Equation~\ref{eq:para}, we finally have
\begin{linenomath*}
	\begin{equation}
		{\hat A_i} = \left\{ {
		\begin{array}{*{20}{c}}
			\begin{aligned}
				&{{A_0}},\\
				&{A_i^b},\\
				&{{A_0} - A_i^s},
			\end{aligned}
		\end{array}
		\begin{array}{*{20}{c}}
			\begin{aligned}
				&{i \in {{\mathcal P}^a}}\\
				&{i \in {{\mathcal P}^b} \cup {{\mathcal P}^s}}\\
				&{i \in {{\mathcal P}^{a \wedge s}}}
			\end{aligned}
		\end{array}} \right..
		\label{eq:Ahat}
	\end{equation}
\end{linenomath*}
Note that ghost air particles are no longer required in calculating $\hat A_i$.

\begin{figure}[t]
	\centering
	\includegraphics[width=1.0\linewidth]{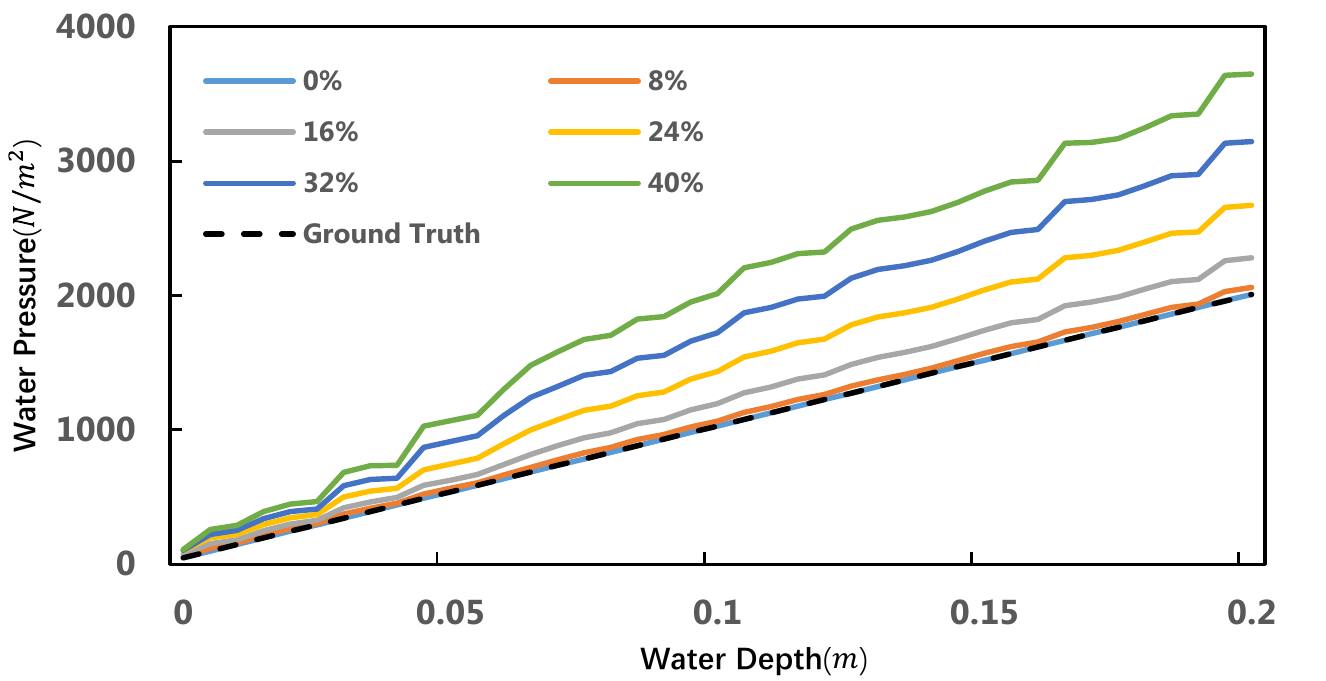}
	\caption{\label{fig:perturbation} A hydrostatic water test solved on a uniformly distributed particles with four neighbors. By introducing a random particle shifting, whose maximum magnitude is $\varepsilon d_0$, to the position field, the simulation accuracy largely decreases as the value of $\varepsilon$ increases from $0\%$ to $40\%$. }
\end{figure}

\subsection{Correcting the source term}
If the particle distribution is regular enough, the following condition should be satisfied for all particles
\begin{linenomath*}
	\begin{equation}
		\mathcal{C}_i = \frac{1}{{{\hat{\alpha} _i}}}  \sum\limits_{j} \frac{{\bf{n}}_{ij}}{{r_{ij}}} {\omega _{ij}}  = 0.
		\label{eq:grad0}
	\end{equation}
\end{linenomath*}
\highlight{By invoking the relationship ${\hat{\alpha} _i} = {\hat{\alpha} _j}$ for a regular particle distribution, an equivalent relationship is obtained as follows
\begin{linenomath*}
	\begin{equation}
		\sum\limits_{j} \left(\frac{1}{{{\hat{\alpha} _i}}} + \frac{1}{{{\hat{\alpha} _j}}} \right) {\mathbf{v}_i^*} \cdot \frac{{\bf{n}}_{ij}}{{r_{ij}}} {\omega _{ij}}  = 0.
		\label{eq:gradv}
	\end{equation}
\end{linenomath*}
Subtracting Equation~\ref{eq:gradv} from the right hand side of Equation~\ref{eq:p2}, the source term becomes
\begin{linenomath*}
	\begin{equation}
		\mathcal{D}_i = \frac{1}{\Delta t} \sum\limits_j \left( {\frac{1}{{{\hat{\alpha} _i}}}+\frac{1}{{{\hat{\alpha} _j}}}} \right) {\left( { \frac{{\mathbf{v}_j^*} - {\mathbf{v}_i^*}}{2}} \right)} \cdot \frac{{{\mathbf{n}_{ij}}}}{{{r_{ij}}}}{\omega _{ij}}
		\label{eq:Di}
	\end{equation}
\end{linenomath*}}
The advantage of applying Equation~\ref{eq:Di} to compute the divergence of velocity is that this new formulation guarantees a zeroth-order accuracy for arbitrary particle distributions while the previous one will fail to correctly compute the divergence for a constant velocity field.
To integrate boundary conditions for $\mathcal{D}_i$, we assume the velocity of ghost air neighbors is equal to ${\bf{v}}_i$.
\highlight{By additionally invoking the solid wall boundary condition defined in Equation~\ref{eq:wall} and assume $\hat \alpha _{j^s} = \hat \alpha_i$, the source term $\mathcal{D}_i$ for all fluid particles can be expressed as
\begin{linenomath*}
	\begin{equation}
		\begin{array}{l}
			\begin{aligned}
			{{\mathcal D}_i} = &\frac{1}{{\Delta t}}\sum\limits_{{j^b}} {\left( {\frac{1}{{{{\hat \alpha }_i}}} + \frac{1}{{{{\hat \alpha }_j}}}} \right)\left( {\frac{{{\bf{v}}_j^* - {\bf{v}}_i^*}}{2}} \right)}  \cdot {{\bf{n}}_{ij}}\frac{{{\omega _{ij}}}}{{{r_{ij}}}}\\
			&{\kern 5pt}+ \frac{1}{{\Delta t}} \left\{ {\begin{array}{*{20}{l}}
				\begin{aligned}
					&{0,{\kern 76pt}i \in {{\mathcal P}^a} \cup {{\mathcal P}^b}}\\
					&{\sum\limits_{{j^s}} {\frac{{2\Delta {\bf{v}}_{i{j}}^*}}{{{{\hat \alpha }_i}}}}  \cdot {{\bf{n}}_{ij}}\frac{{{\omega _{ij}}}}{{{r_{ij}}}},{\kern 11pt}i \in {{\mathcal P}^s} \cup {{\mathcal P}^{a \wedge s}}}
				\end{aligned}	
		\end{array}} \right.
		\end{aligned}	
		\end{array}.
		\label{eq:div}
	\end{equation}
\end{linenomath*}}

\subsection{Particle Shifting}
The condition in Equation~\ref{eq:grad0} will be destroyed when particles move anisotropically.
To investigate how an irregular particle distribution can affect the fluid evolution, let us consider solving a hydrostatic water test on uniformly distributed particles where only four nearest particles are stored as neighbors.
We then introduce a random shift $\delta {\bf{x}}_i$, whose maximum shifting distance is defined as $\varepsilon d_0$, to the position field.
By setting $\varepsilon$ to different values, a set of pressure fields at $t=0$ can be solved and plotted at Figure~\ref{fig:perturbation}.
From the comparison, it can be noted that the simulation accuracy significantly decreases as the magnitude of random particle shifting increases.
Therefore, it is necessary to maintain the regularity of particle distribution with particle shifting.
\highlight{Motivated by the finite particle volume method~\cite{Nestor:2008:MBP}, Xu et al.~\cite{Xu:2009:ASI} first applied a particle shifting algorithm to avoid non-uniform particle distributions in ISPH, yet it suffers from instabilities for flows near the free surface.
Later works~\cite{Lind:2012:ISP,Skillen:2013:ISP} have stabilized the particle shifting algorithm by proposing to govern the magnitude and direction of the position shift according to Fick's law.
The idea is to shift particles' positions slightly from regions of high particle concentration to regions of low concentration.}
\highlight{However, it still results in numerical inconsistencies probably due to inaccuracies in calculation of normal vectors at free-surface and implementation of inconsistent particle shifting displacement equations~\cite{Khayyer:2017:CSA}.
Since flows near free-surface are usually related to surface tension effects, we are motivated to combine the particle shifting with some kind of surface tension model to form a new particle shifting algorithm that can both regularize particle distributions and capture surface tension effects.}

Inspired by the Helmholtz free energy functional~\cite{Cahn:1958:FEN,He:2014:RSS,Yang:2015:FMF}, we propose a new particle shifting algorithm that minimizes the following energy
\begin{equation}
	{\mathcal{F}}_i = \frac{1}{2}\frac{\left\| {{{\bf x}_i} - {\bf x}_i^*} \right\|^2}{d_0^2} + f\left( c_i \right) + \frac{\kappa}{2} \left\| \nabla _i c \right\|^2
	\label{eq:Helmholtz}
\end{equation}
by treating all particles as having a volume of 1, where $c$ is the concentration variable, $d_0$ is the sampling distance, $\kappa$ is a squared gradient energy coefficient.
The first term $\frac{1}{2}{\left\| {{{\bf x}_i} - {\bf x}_i^*} \right\|^2}/{d_0^2}$ can be viewed as a momentum potential~\cite{Bouaziz:2014:PDF}, which is included to guarantee the particle movement is as small as possible.
\highlight{The second term $f(c_i)$ represents the bulk energy density, which guarantees the total volume is preserved.}
Its exact formulation is defined as
\begin{equation}
	f(c_i) = \frac{\lambda }{4}{\left( {\frac{{c_i^2}}{{c_0^2}} - 1} \right)^2},
	\label{eq:bulk}
\end{equation}
where $\lambda$ is the bulk energy coefficient and $c_0$ is a reference value for $c_i$.
By taking a weighting function $W(r)$ that meets $\partial W / \partial r = \omega / r$ and defining $c_i$ as
\begin{equation}
	{c_i} = \frac{1}{{{\hat \alpha _i}}}\sum\limits_j {{W_{ij}}},
	\label{eq:ci}
\end{equation}
we can finally compute $\nabla _i c$ as
\begin{equation}
	\nabla _i c = \frac{1}{{{\hat{\alpha} _i}}}  \sum\limits_{j} \frac{{\bf{n}}_{ij}}{{r_{ij}}} {\omega _{ij}}.
\end{equation}
Minimizing the squared gradient energy density is equivalent to imposing the condition in Equation~\ref{eq:grad0} for interior particles.
The advantage of transferring Equation~\ref{eq:grad0} into the energy minimization problem is that the squared gradient energy enables us to capture the correct surface tension effects for boundary particles as well~\cite{He:2014:RSS}.

\begin{figure}[t]
	\centering
	\subfigure[No particle shifting]{\includegraphics[width=0.495\linewidth]{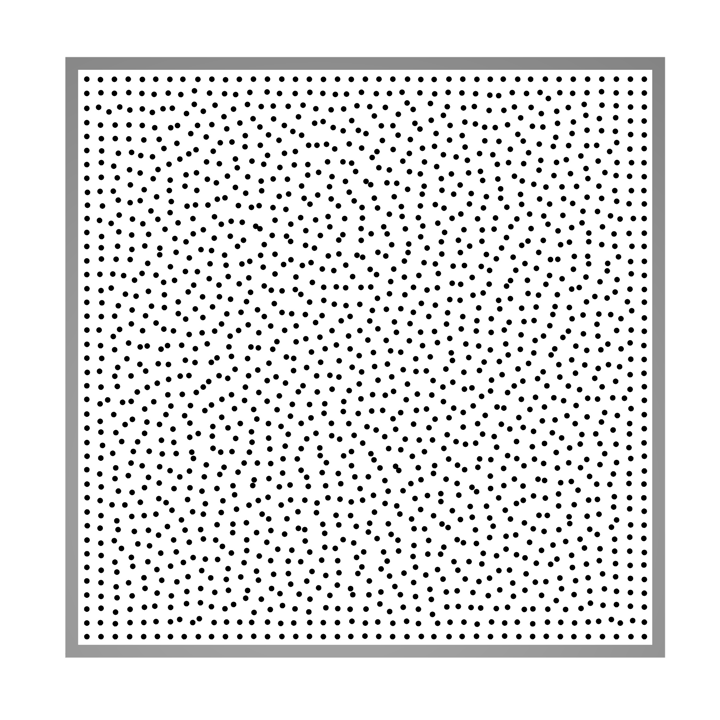}}
	\subfigure[With particle shifting]{\includegraphics[width=0.495\linewidth]{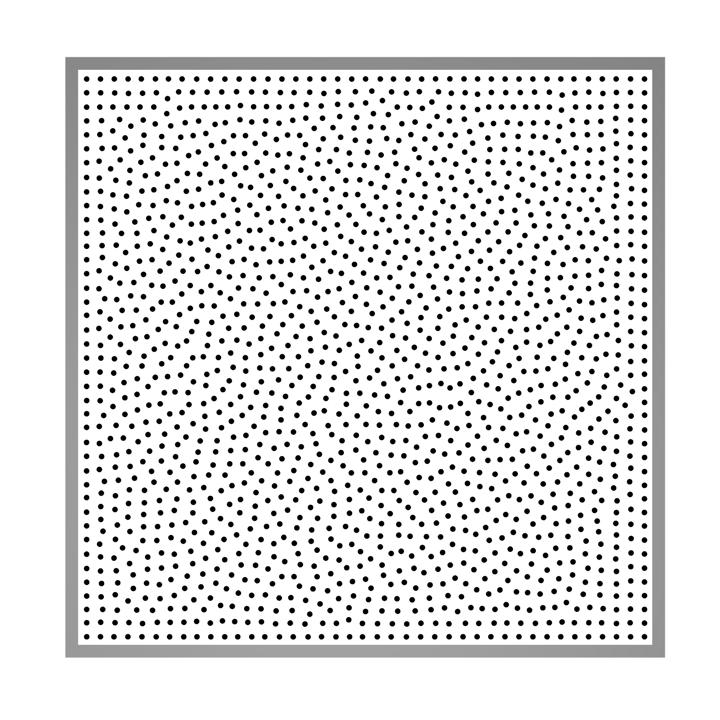}}
	\caption{\label{fig:vortex} Particle distributions in a Taylor–Green vortex at $t = 0.4s$.}
\end{figure}
\begin{figure}[t]
	\centering
	\includegraphics[width=0.325\linewidth]{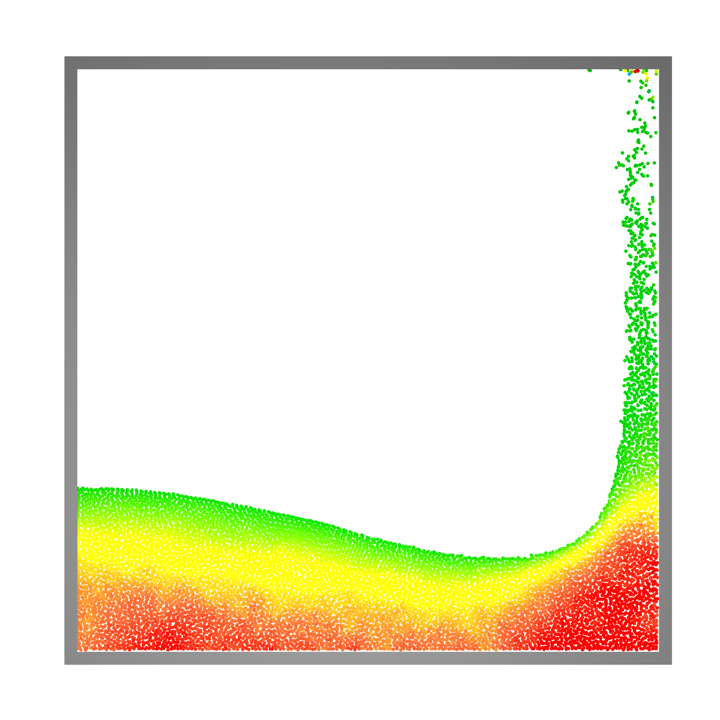}
	\includegraphics[width=0.325\linewidth]{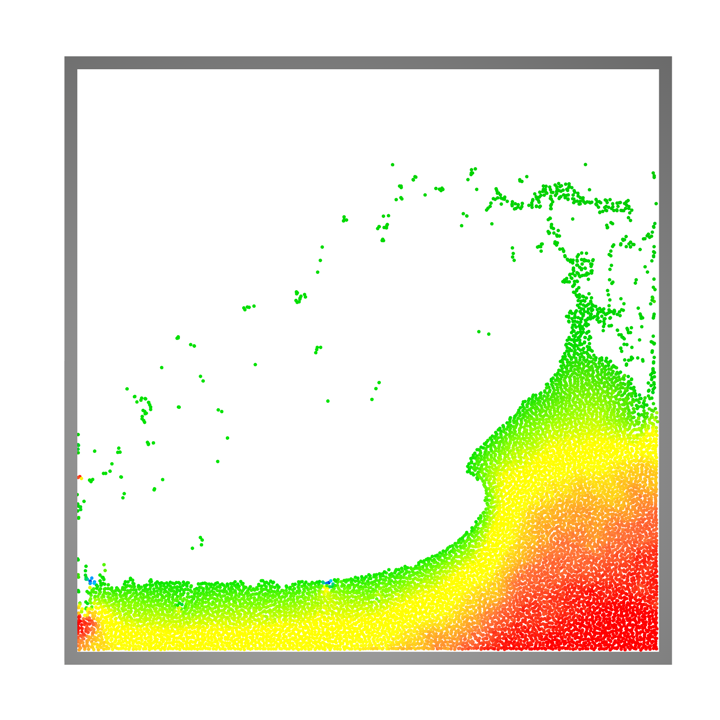}
	\includegraphics[width=0.325\linewidth]{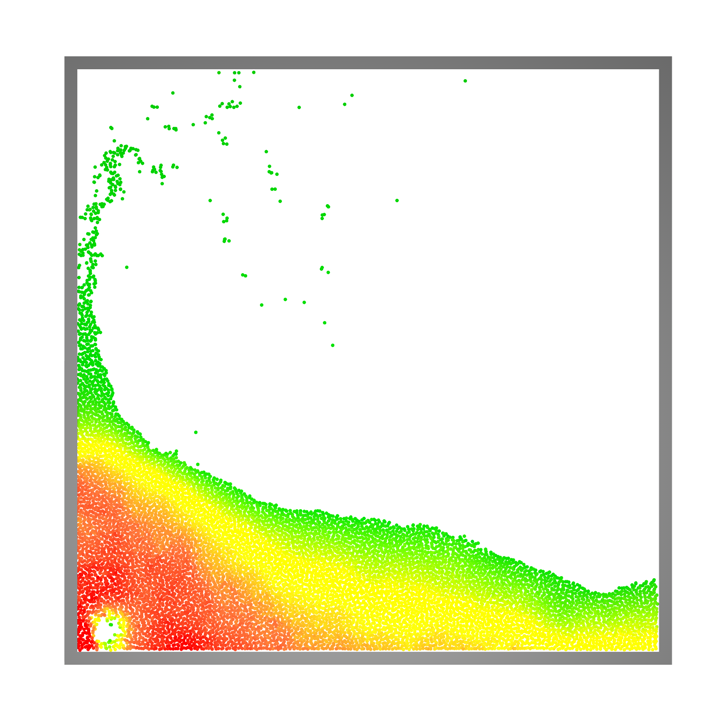}
	\includegraphics[width=0.325\linewidth]{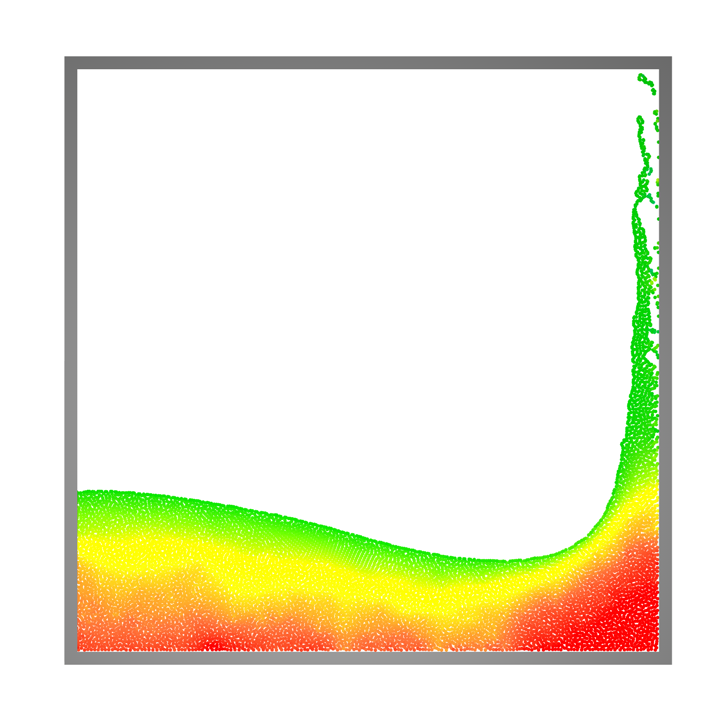}
	\includegraphics[width=0.325\linewidth]{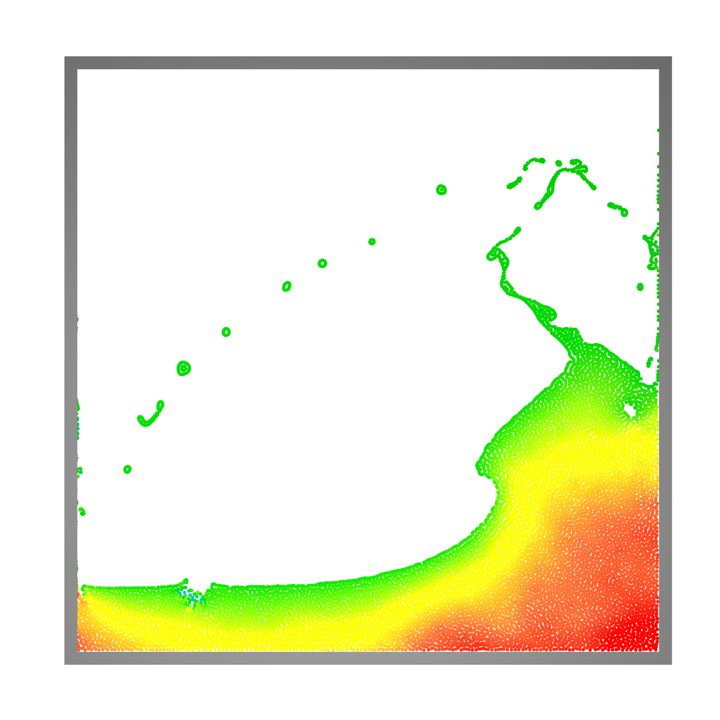}
	\includegraphics[width=0.325\linewidth]{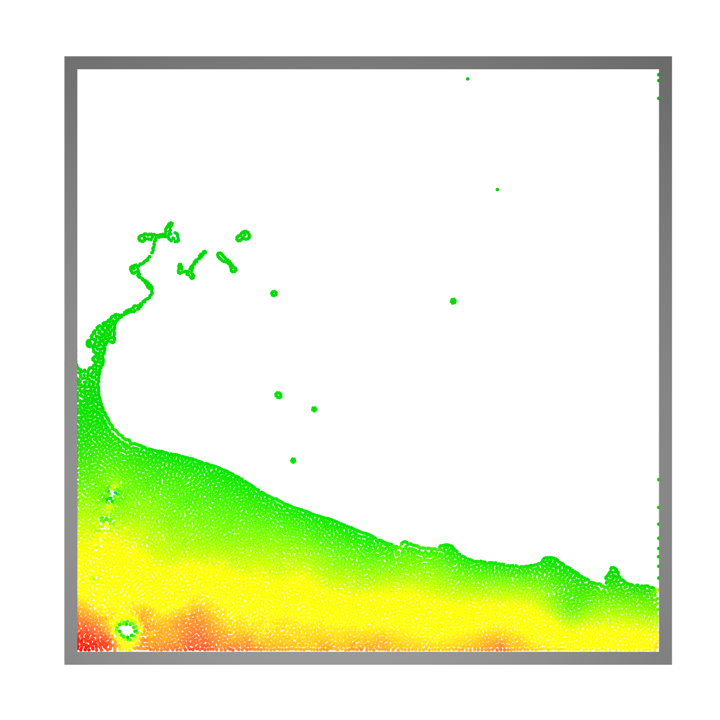}
	\caption{\label{fig:2ddambreak} Side by side comparison of a two-dimensional dambreak example between taking a particle shifting algorithm with surface tension effects (Bottom) and without taking the particle shifting (Top).}
\end{figure}
To solve the optimization problem, we apply a strategy based on Fick’s law of diffusion to regularize the particle distribution iteratively~\cite{Lind:2012:ISP}.
Its purpose is to shift particles from regions of high Helmholtz free energy concentration to regions of low concentration.
At each iteration, the displacement vector for particle $i$ is written as
\begin{equation}
	\delta {\bf x}_i = -\varsigma \nabla _i \mathcal{F}
\end{equation}
where $\nabla _i {\mathcal F}$ is computed by taking the derivative of ${\mathcal F}_i$ with respect to ${\bf x}_i$
\begin{equation}
\nabla _i {\mathcal F} = \frac{{{{\bf{x}}_i} - {\bf{x}}_i^*}}{{d_0^2}} + \lambda \left( {\frac{{c_i^3}}{{c_0^4}} - \frac{{{c_i}}}{{c_0^4}}} \right)\nabla _i {c} + \kappa \left({\Delta _i}c\right){\nabla _i}c .
\end{equation}
Here ${\Delta _i}c$ is defined as
\begin{equation}
	{\Delta _i}c = \frac{1}{{{{\hat \alpha }_i}}}\sum\limits_j {\frac{{\omega _{ij}^{'}}}{{{r_{ij}}}}}.
	\label{eq:lapc}
\end{equation}
$\varsigma$ is a coefficient that controls the distance a particle moves during one iteration.
By noting that
\begin{equation}
	\begin{array}{l}
		\begin{aligned}
		\left\| {\nabla _i {\mathcal F}} \right\| &\le \left\| {\frac{{{{\bf{x}}_i} - {\bf{x}}_i^*}}{{d_0^2}}} \right\| + \lambda \left\| {\left( {\frac{{c_i^3}}{{c_0^4}} - \frac{{{c_i}}}{{c_0^4}}} \right)\nabla {c_i}} \right\| + \kappa \left\| {\left( {{\Delta _i}c} \right){\nabla _i}c} \right\|\\
		 &\le \frac{1}{{{d_0}}} + \frac{\lambda _0 }{{{c_0}}} + \kappa _0\left( {{\Delta _0}c} \right),
		\end{aligned}
	\end{array}
\end{equation}
where ${{\Delta _0}c}$ is a reference value for ${{\Delta _i}c}$, $\lambda _0$ and $\kappa _0$ are the upper limits for $\lambda$ and $\kappa$ which are simply set to 1, we define the coefficient $\varsigma$ to be as
\begin{equation}
	\varsigma = {{{d_0}} \mathord{\left/
 {\vphantom {{{d_0}} {\left[ {\frac{1}{{{d_0}}} + \frac{\lambda }{{{c_0}}} + \kappa \left( {{\Delta _0}c} \right)} \right]}}} \right.
 \kern-\nulldelimiterspace} {\left[ {\frac{1}{{{d_0}}} + \frac{\lambda _0}{{{c_0}}} + \kappa _0 \left( {{\Delta _0}c} \right)} \right]}}.
	\label{eq:varsigma}
\end{equation}
Therefore, an upper limit of $d_0$ is imposed on the particle shifting distance for one iteration.

\begin{figure*}[t]
	\centering
	\includegraphics[width=1.0\linewidth]{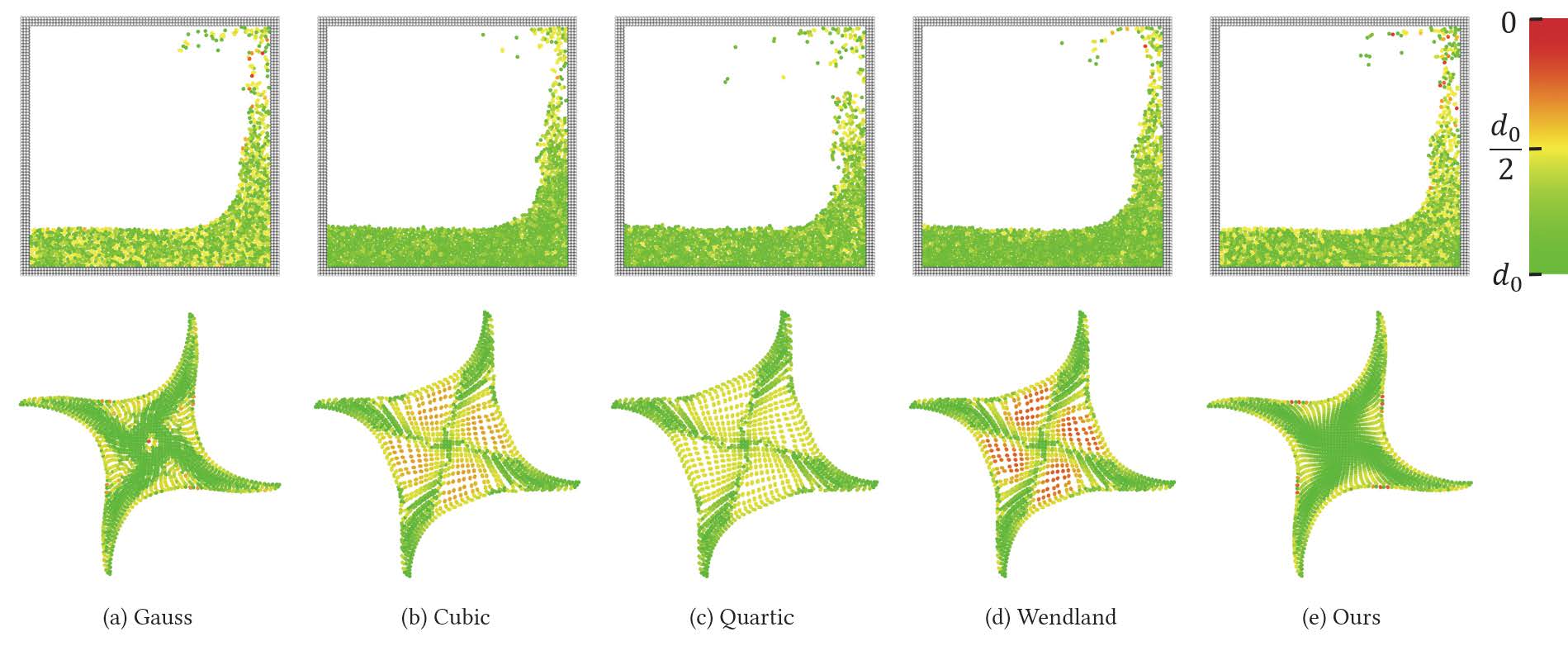}
	\caption{\label{fig:kernelcmp} \highlight{Evaluation of the particle clumping problem with five different kernel functions. Top:  A 2D dambreak is used for evaluation of numerical schemes in suppression of pairing instability; Bottom: A rotating square patch is used as a benchmark test for evaluation of numerical schemes in suppression of tensile instability.}}
\end{figure*}

Figure~\ref{fig:vortex} demonstrates the effectiveness of our particle shifting algorithm in regularizing the distribution for interior particles.
For this example, $\kappa$ is simply set to $0$ as there is no free surface boundary.
Besides, since our particle shifting algorithm is motivated by the Helmholtz free energy functional, we can also introduce surface tension effects by independently adjusting the value of $\kappa$, see Figure~\ref{fig:2ddambreak} for a demonstration.

\section{Addressing the Particle Clumping Problem}
There are two typical models to compute the pressure force under the SPH framework: the symmetric repulsive pressure gradient model~\cite{Ihmsen:2014:IIS,Yang:2015:FMF,Winchenbach:2017:ICA,Gissler:2019:ISP} and the Taylor-series consistent pressure gradient model~\cite{He:2012:SMS,Yang:2016:ESS}.
The symmetric repulsive pressure gradient model has been widely used in SPH due to its superior stability features~\cite{Shao:2003:ISM,Khayyer:2008:CIS}.
However, this model is more sensitive to the tensile instability problem~\cite{Gotoh:2016:CAF}.
Therefore, we apply the Taylor-series consistent pressure gradient model to compute the pressure force as follows
\begin{linenomath*}
	\begin{equation}
		{\mathbf{F}^p_i} = \frac{1}{\rho _0}
		{\sum\limits_{j} \left( \frac{1}{{{\hat{\alpha} _i}}} + \frac{1}{{{\hat{\alpha} _j}}} \right)\left( {{p_j} - {p_i}} \right){{{\bf{n}}_{ij}}}\frac{{{\omega _{ij}}}}{{{r_{i{j}}}}},}
		\label{eq:tf}
	\end{equation}
\end{linenomath*}

Next, we will discuss how to improve the Taylor-series consistent pressure gradient model to avoid pairing instability.
\highlight{Assume $W$ is a kernel function that is commonly used in SPH.
According to the Swegle’s condition of instability, a sufficient criterion for unstable growth in compressive regime is $W {''}(r) < 0$ ( because the compressive stress is assumed to be negative in~\cite{Swegle:1995:SPH} ),
where $W {''}$ represents the second derivative of the kernel function.
Invoking the relationship between $\omega$ and $W$ in Equation~\ref{eq:w_relationship}, an equivalent sufficient criterion for unstable growth in terms of $\omega$ should be as follows
\begin{equation}
\Omega_{\omega}(r)=\frac{\omega}{{{r^2}}} - \frac{{\omega{'}}}{r} < 0.
\label{eq:criterion}
\end{equation}
That is say, in order to avoid pairing instability, it is better for us to select a kernel function satisfying $\Omega_{\omega}(r) > 0$.}

\highlight{Before proceeding with the choice of kernel function, it should be pointed out that Dehnen and Aly~\cite{Dehnen:2012:ICS} disproved Swegle’s statement by showing that the Wendland functions (whose second derivative cannot be strictly larger than 0) can avoid the pairing instability for all scale of neighbors in WCSPH.
Unfortunately, as we apply the Wendland function ( e.g., by setting $W(r)=\left(1-\frac{r}{h}\right)^3\left(1+3\frac{r}{h}\right)$ and $\omega(r)=-rW'=12\frac{r^2}{h^2}\left(1-\frac{r}{h}\right)^2$ ) within our incompressible fluid solver, severe pairing instability is observed for a 2D dambreak test.
Therefore, we still use Equation~\ref{eq:criterion} as a guidance for selecting the appropriate kernel function.
Since the sufficient criterion for unstable growth is now given in terms of $\omega$, we actually have more freedom in selecting the kernel function, e.g., we can simply set $\omega$ rather than $W$ to be a Wendland function.}
However, this also introduces another numerical problem involving the terms $\omega / r$ and $\omega / {r^2}$, since both the value of $\omega / r$ and $\omega / {r^2}$ in Equation~\ref{eq:p2} could be extremely large when two neighboring particles are too close to each other, resulting in simulation failure.
To solve this problem, we propose to correct the kernel function as follows
\begin{linenomath*}
	\begin{equation}
		\frac{ \omega }{r} = \left\{ {\begin{array}{*{20}{c}}
				\begin{aligned}
					{\frac{ \omega(r) }{r},}{\kern 5pt}&{r > \delta }\\
					{\frac{ \omega(r) }{\delta },}{\kern 5pt}&{r \le \delta }
				\end{aligned}
		\end{array}} \right.,
		{\kern 10pt}
		\frac{ \omega }{r^2}= \left\{ {\begin{array}{*{20}{c}}
				\begin{aligned}
					{\frac{ \omega(r) }{{{r^2}}},}{\kern 5pt}&{r > \delta }\\
					{\frac{ \omega(r) }{{{\delta ^2}}},}{\kern 5pt}&{r \le \delta }
				\end{aligned}
		\end{array}} \right.
	\end{equation}
\end{linenomath*}
where $\delta$ is a threshold used to prevent $\omega / r$ and $\omega / r^2$ from generating too large values.
Unless stated, we will always set $\delta$ to be the particle sampling distance for all examples.

\begin{figure}[t]
	\centering
	\includegraphics[width=1.0\linewidth]{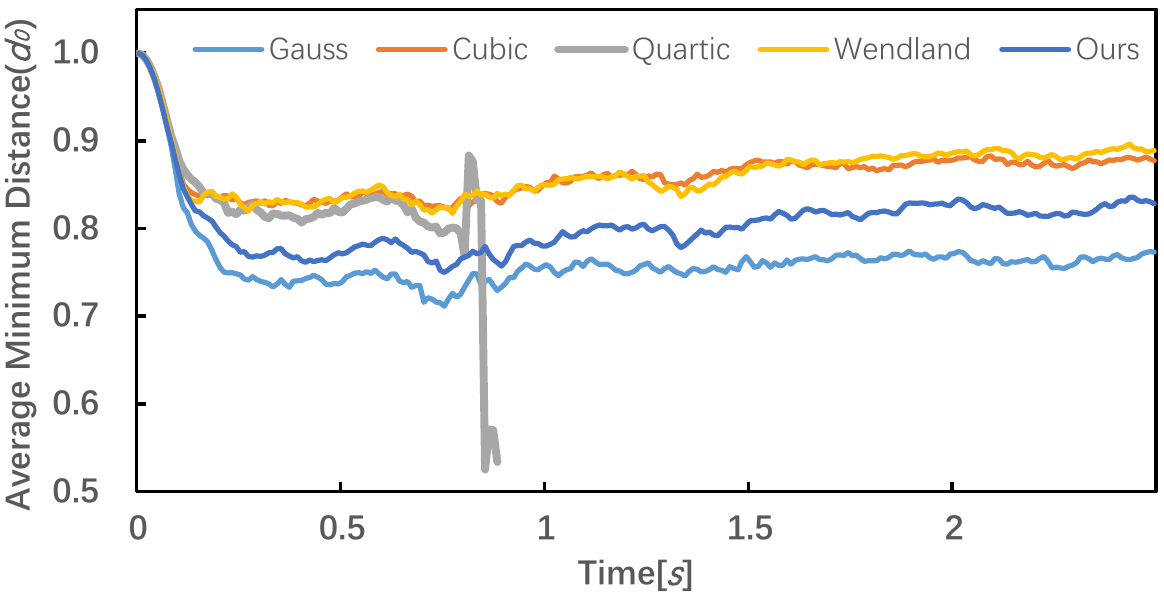}
	\caption{\label{fig:distance} \highlight{Time history of the average of $\bar{d}$ for all dambreak simulations in Figure~\ref{fig:kernelcmp}. Note the simulation with the Quartic kernel function fails at around $t=0.8s$.} }
\end{figure}

\highlight{Motivated by the kernel function commonly used in MPS~\cite{Koshizuka:1996:MPS}, we propose to use the following kernel function}
\begin{equation}
	\omega  = \left\{ {\begin{array}{*{20}{c}}
		\begin{aligned}
			&{1 - \frac{{{r^4}}}{{{h^4}}},{\kern 15pt}r < h}\\
			&{0,{\kern 36pt}r \ge h}
\end{aligned}
	\end{array}} \right..
\label{eq:kernel}
\end{equation}
\highlight{It can be easily verified $\Omega_{\omega}(t) > 0$ stands for all $r \in [0, h]$.
To study its performance, Figure~\ref{fig:kernelcmp} tests two examples by selecting a set of different kernel functions.
The rotating square patch is commonly used as a benchmark test for evaluation of numerical schemes in suppression of tensile instability while the dambreak is used for evaluation of pairing instability.
In order to quantify simulation outcomes, we compute an average of the minimum distance of each particle to its neighbors as
\begin{equation}
\bar d = \frac{1}{N}\sum\limits_i {\mathop {\min }\limits_{i \ne j} \left\| {{\mathbf{x}_i} - {\mathbf{x}_j}} \right\|}.
\end{equation}
Figure~\ref{fig:distance} plots the time history of $\bar d$ for all dambreak simulations.
From the comparison, it can be noticed both the Cubic and Wendland functions perform best in avoiding the pairing instability.
However, in testing the rotation square example, simulations with the Cubic and Wendland functions break up much earlier than the one with our kernel function, which indicates our kernel function is better in suppressing the tensile instability.
Since this work is focused on creating viscous fingering structures in fluids, we will always select Equation~\ref{eq:kernel} as the kernel function in the following discussions.
However, if researchers are interested in simulating examples mainly under compressive stress states, we suggest to use the Cubic or Wendland functions.}

\begin{algorithm}[t]
	\caption{Variational Staggered Incompressible SPH}
	\label{alg:alg}
	\begin{algorithmic}[1]
		\State  Precompute ${A_0}$, ${\alpha}_0$, $\beta_0$, $\nabla _0 c$ and ${{\Delta _0}c}$;
		\State  Initialize normal vectors for all ghost solid particles;
		\While{$t < t_{stop}$}
		\State $\Delta t \gets CFL(\mathbf{v}_i^t)$;
		\ForAll{fluid particle $i$}
		\State   ${\mathcal N}_i \gets$ Find neighbors;
		\EndFor
		\ForAll{fluid particle $i$}
		\State   $\mathbf{v}_i^* \gets \mathbf{v}_i^t + \Delta t ( \mathbf{F}^{\upsilon} + \mathbf{F}^{ext} )$;
		\State   $\mathbf{x}_i^* \gets \mathbf{x}_i^t + \Delta t \mathbf{v}_i^*$;
		\EndFor
		\State  Run the particle shifting algorithm;
        \ForAll{fluid particle $i$}
		\State Compute $A_i$ and $\alpha _i$;
        \EndFor
		\State Categorize fluid particles into four subsets;
		\ForAll{fluid particle $i$}
        \State  Initialize the particle pressure $p_i^0$;
		\State Compute ${\mathcal L} _i^0$ according to Equation~\ref{eq:L2} and~\ref{eq:Ahat};
		\State Compute $\mathcal{D}_i^0$ according to Equation~\ref{eq:div};
        \State Compute the residual $q_i^0 = {\mathcal L}_i^0 - {\mathcal D}_i^0$;
        \State Set $y_i^0 = q_i^0$;
        \EndFor
        \State  Set the iteration number $k = 0$;
        \While  {${\eta ^k} > \eta$}
        \ForAll{fluid particle $i$}
        \State  Compute $\tilde {\mathcal L}_i^k$ in the same way as ${\mathcal L}_i^k$ except $p_i^k$ is replaced with $y_i^k$;
        \EndFor
        \State  Compute ${\beta ^k} = {{\sum\nolimits_i \left( {q_i^k \cdot q_i^k} \right) } / {\sum\nolimits_i \left( {{y_i^k} \cdot \tilde {\mathcal L}_i^k} \right) }}$;
        \ForAll{fluid particle $i$}
        \State  Update the pressure $p_i^{k + 1} = p_i^k + {\beta ^k}r_i^k$;
        \State  Update the residual $q_i^{k + 1} = q_i^k - {\beta ^k}{{\tilde {\mathcal L}}_i}$;
        \EndFor
        \State  Compute ${\gamma ^k} = {{\sum\nolimits_i \left( {q_i^{k + 1} \cdot q_i^{k + 1}} \right) } / {\sum\nolimits_i \left( {q_i^k \cdot q_i^k} \right) }}$;
        \ForAll{fluid particle $i$}
        \State  Update $y_i^{k + 1} = q_i^{k + 1} + {\gamma ^k}y_i^k$;
        \EndFor
        \EndWhile
		\ForAll{fluid particle $i$}
		\State Compute $\mathbf{F}^p_i$ according to Equation~\ref{eq:pffinal} with $p_i^{k+1}$;
		\State $\mathbf{v}_i^{t+\Delta t} \gets \mathbf{v}_i^* + \Delta t \mathbf{F}^p_i$;
		\EndFor
		\EndWhile
	\end{algorithmic}
\end{algorithm}
Till now, \highlight{the last problem} that has not been addressed is how to integrate all boundary conditions when calculating the pressure force.
By invoking the free surface boundary condition $p_{j^s} = 0$, solid wall boundary condition defined in Equation~\ref{eq:wall} and the condition in Equation~\ref{eq:grad0}, we can reformulate the pressure force according to different types of fluid particles
\begin{linenomath*}
	\begin{equation}
		{\kern -5pt}{\bf{F}}^p_i = \frac{\beta_0}{{{\rho _0}}}\left\{ {\begin{array}{*{20}{c}}
			\begin{aligned}
			&{\sum\limits_{{j^b}} {\left( {\frac{1}{{{{\hat \alpha }_i}}} + \frac{1}{{{{\hat \alpha }_j}}}} \right)} {\kern 3pt} {p_j}{{\bf{n}}_{ij}}\frac{{{\omega _{ij}}}}{{{r_{ij}}}},{\kern 49pt}i \in {P^a}}\\
			&{\sum\limits_{{j^b}} {\left( {\frac{1}{{{{\hat \alpha }_i}}} + \frac{1}{{{{\hat \alpha }_j}}}} \right)} \left( {{p_j} - {p_i}} \right){{\bf{n}}_{ij}}\frac{{{\omega _{ij}}}}{{{r_{ij}}}},{\kern 24pt}i \in {P^b}}\\
			&{\sum\limits_{{j^b}} {\left( {\frac{1}{{{{\hat \alpha }_i}}} + \frac{1}{{{{\hat \alpha }_j}}}} \right) {\kern 3pt}{p_j}{{\bf{n}}_{ij}}\frac{{{\omega _{ij}}}}{{{r_{ij}}}}} +  {\Lambda _i^a} + {\Lambda _i^s},{\kern 6pt}i \in {P^{a \wedge s}}}\\
			&{\sum\limits_{{j^b}} {\left( {\frac{1}{{{{\hat \alpha }_i}}} + \frac{1}{{{{\hat \alpha }_j}}}} \right)} \left( {{p_j} - {p_i}} \right){{\bf{n}}_{ij}}\frac{{{\omega _{ij}}}}{{{r_{ij}}}} + {\Lambda _i^s},{\kern 3pt}i \in {P^s}}
			\end{aligned}	
		\end{array}} \right.
		\label{eq:pffinal}
	\end{equation}
\end{linenomath*}
, where ${\Lambda _i^a}$ and ${\Lambda _i^s}$ are two sources originated from ghost air neighbors and ghost solid neighbors, respectively.
Their formulations are written as
\begin{equation}
	\begin{array}{l}
		\begin{aligned}
		&\Lambda _i^a = \sum\limits_{{j^b}} {\left( {\frac{1}{{{{\hat \alpha }_i}}} + \frac{1}{{{{\hat \alpha }_j}}}} \right){p_i}{{\bf{n}}_{ij}}\frac{{{\omega _{ij}}}}{{{r_{ij}}}}} \\
		&\Lambda _i^s = \sum\limits_{{j^s}} {\frac{2}{{{{\hat \alpha }_i}}}{\rm proj}{_{{{\bf{n}}_{ij^s}}}}(\Delta {\bf{v}}_{i{j^s}}^*){\omega _{ij}}}
		\end{aligned}
	\end{array}.
	\label{eq:psource}
\end{equation}
$\beta_0$ is a constant correction factor for the pressure force because the Taylor-series consistent pressure gradient model defined Equation~\ref{eq:tf} does not have a first-order accuracy.
To reach a first-order accuracy, we first set $\beta_0$ to be 1 and initialize a linear pressure field, e.g., $p = x$, at the beginning of simulation.
Then, we scale the value of $\beta_0$ to make the pressure force match its real value.
The reason we do not apply a first-order Taylor-series pressure gradient model is because higher order models are more sensitive to the particle distribution, especially for boundary particles that suffer the particle deficiency problem~\cite{Fang:2009:ISM}.

\section{Results and Discussions}
\begin{figure}[t]
	\centering
	\subfigure[Color-coded pressure fields at $t = 1.5s$.]{\includegraphics[width=1.0\linewidth]{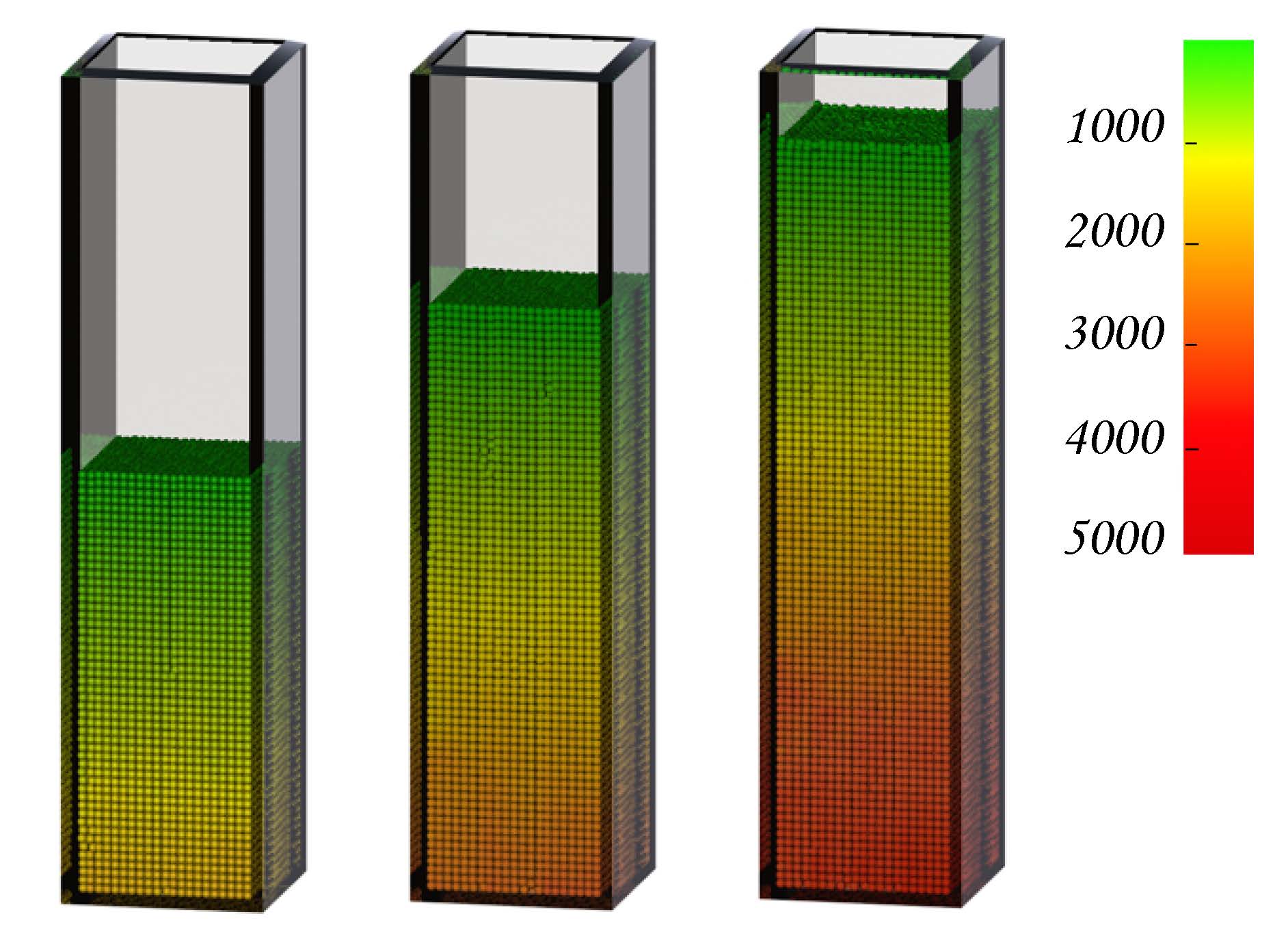}}
	\subfigure[Time history of the water pressure at bottom.]{\includegraphics[width=1.0\linewidth]{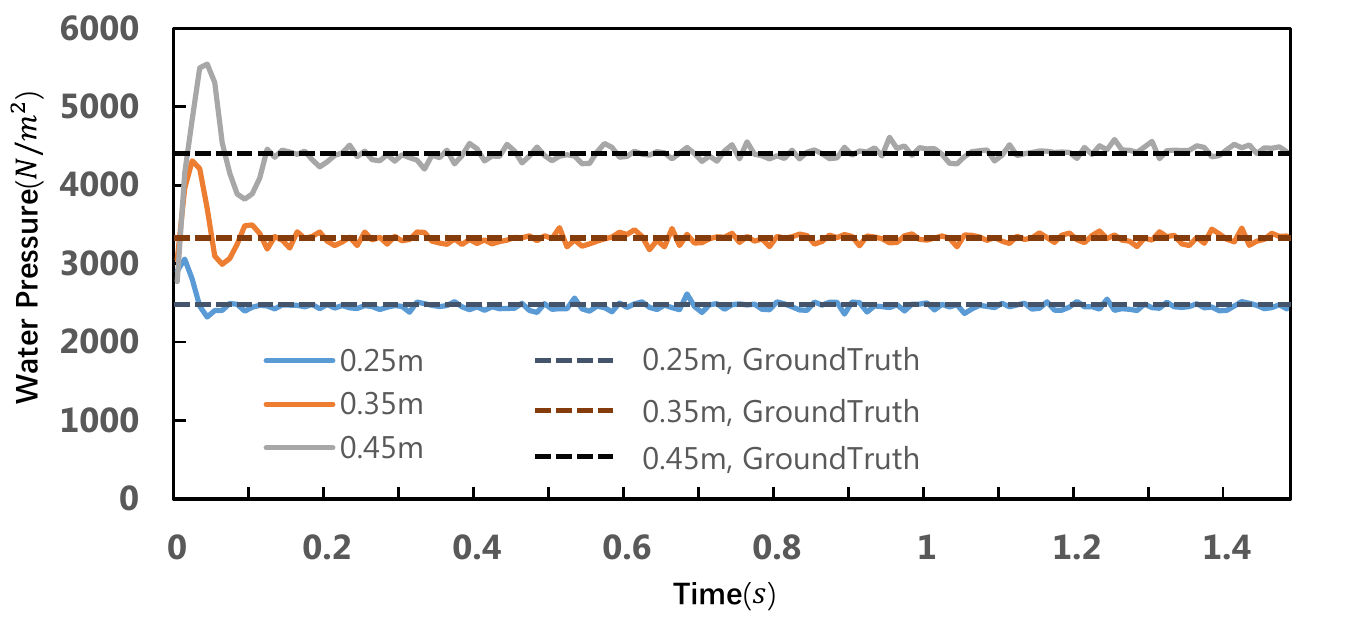}}
	\vspace{-0.1in}
	\caption{\label{fig:static} Three hydrostatic water tests with different heights. }
\end{figure}
\begin{figure*}[t]
	\centering
	\includegraphics[width=1.0\linewidth]{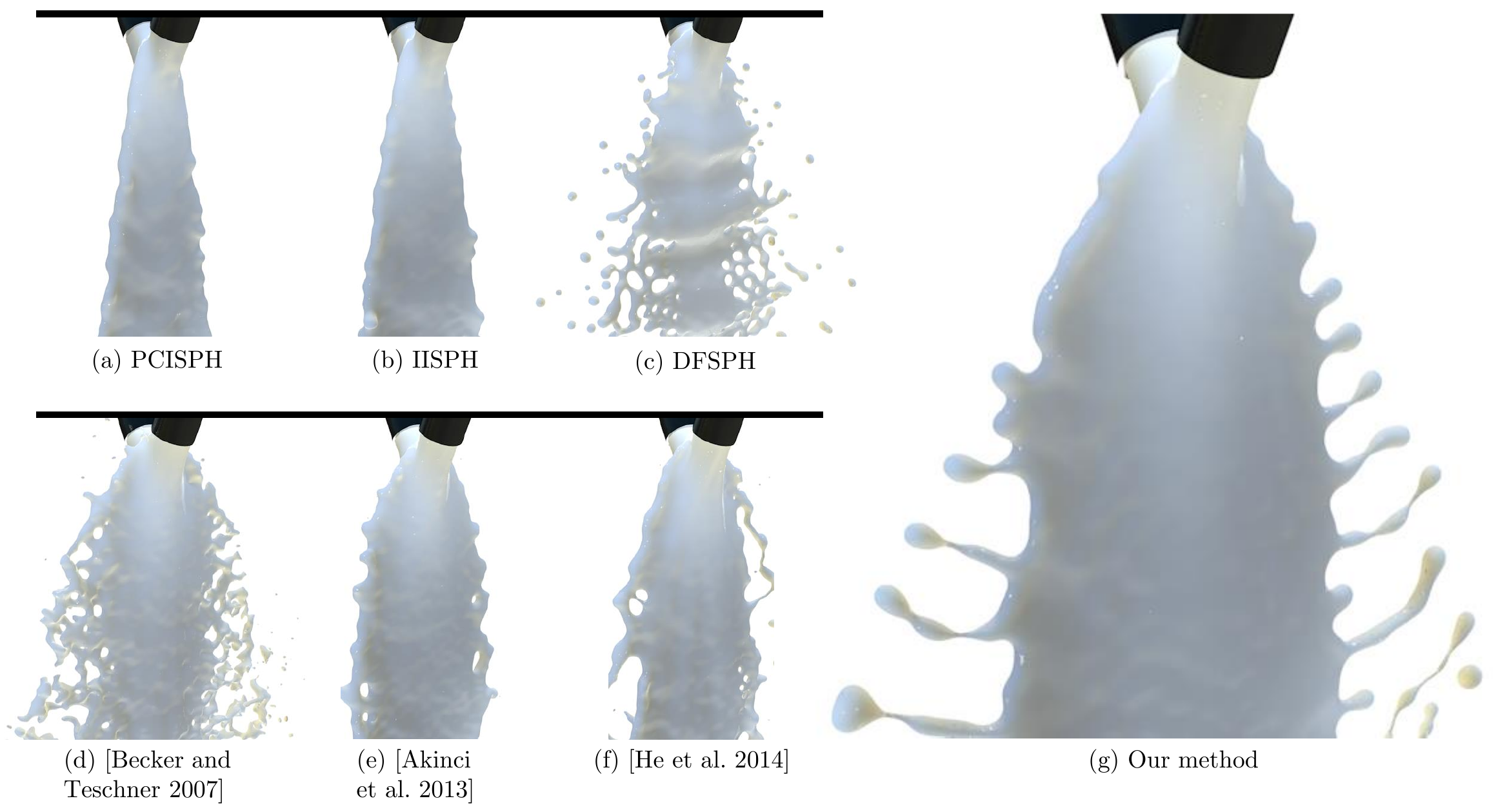}
	\vspace{-0.1in}
	\caption{\label{fig:fishbonecmp} Comparison of two impinging jets simulated with different incompressibility solvers and surface tension models, the inlet velocity is initialized to be $6.5m/s$ for all simulations. }
\end{figure*}

We implement our method with CUDA and run all examples on an NIVDIA Geforce GTX 1060 graphics card.
Algorithm 1 outlines an overview of our method.
Before the simulation starts, we precompute all reference values, including $A_0$, ${\alpha} _0$, $\beta_0$, $\nabla _0 c$ and $\Delta _0 c$, from a prototype particle with full neighbors.
For each time step, we use a prediction-correction scheme similar to the two-step projection method~\cite{Shao:2003:ISM},
\highlight{where the divergence-free condition $\nabla \cdot \mathbf{v} = 0$ is temporarily not fulfilled at the prediction step by advecting particles forward (lines 7$\sim$10) and is later imposed at the correction step by solving a pressure Poisson equation (lines 14$\sim$33).}
We compute an XSPH artificial viscous force $\bf{F}^\upsilon$ following~\cite{Monaghan:1989:PPP,Schechter:2012:GSA} to stabilize inviscid flows.
To avoid cumulative density errors, we additionally add an error compensating source (ECS) to the source term following Khayyer and Gotoh~\cite{Khayyer:2011:ESA}.
\highlight{In our work, the error compensating source can be easily computed as follows
\begin{equation}
ECS_i^{t + \Delta t} = \left| {\frac{{\rho _i^t - {\rho _0}}}{{{\rho _0}}}} \right|{\mathcal D}_i^t + \left| {{\mathcal D}_i^t} \right|\left( {\frac{{\rho _i^t - {\rho _0}}}{{{\rho _0}}}} \right).
\end{equation}
Please also note $ECS_i^{t + \Delta t}$ is only added to particles whose density $\rho _i^t$ is larger than $\rho _0$.}
If not specified, the smooth length $h$ is always set to $h=2.5d_0$.
The time step size is limited by the Courant condition~\cite{Shao:2003:ISM}.
Surface meshes are reconstructed with the particle skinning method~\cite{Bhatacharya:2011:LMS}.
\highlight{The open source code SPlisHSPlasH$\footnote{https://github.com/InteractiveComputerGraphics/SPlisHSPlasH}$ has also been applied for the comparison in Figure~\ref{fig:fishbonecmp}.}

\begin{figure}[t]
	\centering
	\includegraphics[width=1.0\linewidth]{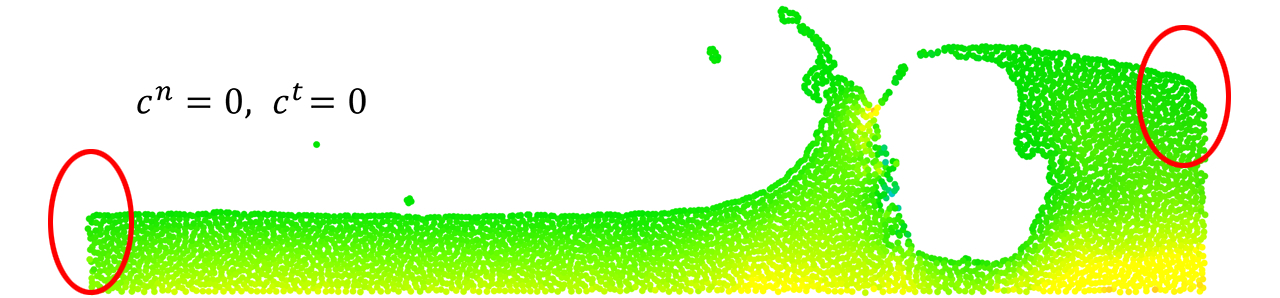}
	\includegraphics[width=1.0\linewidth]{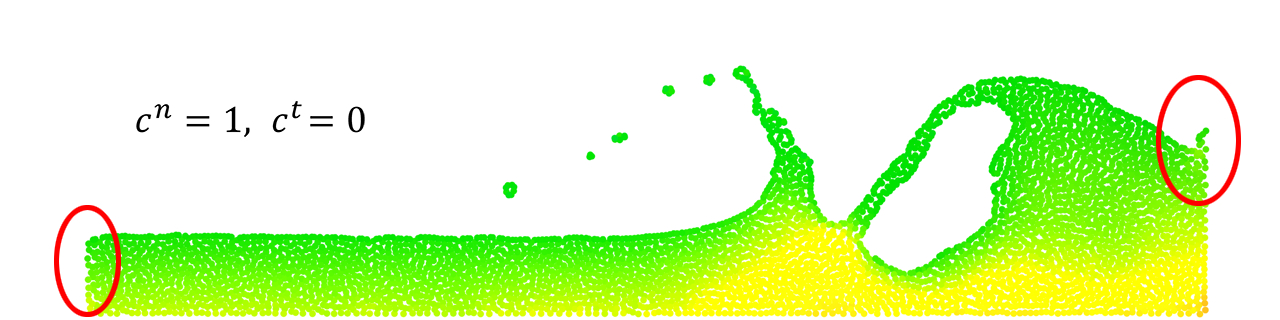}
	\includegraphics[width=1.0\linewidth]{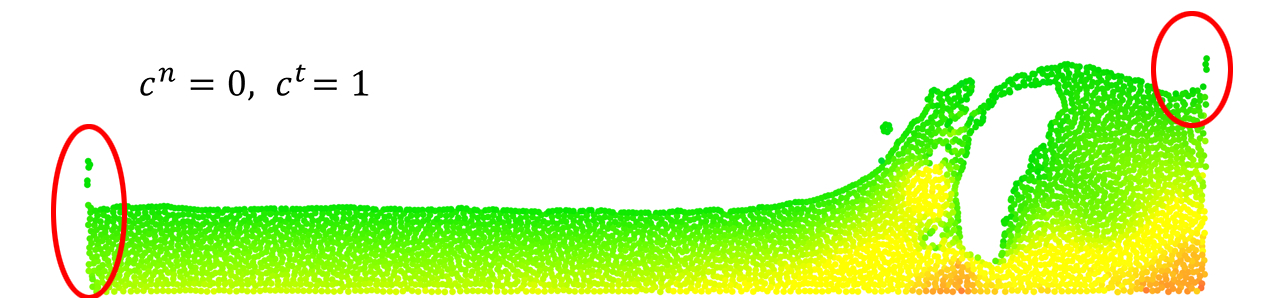}
	\includegraphics[width=1.0\linewidth]{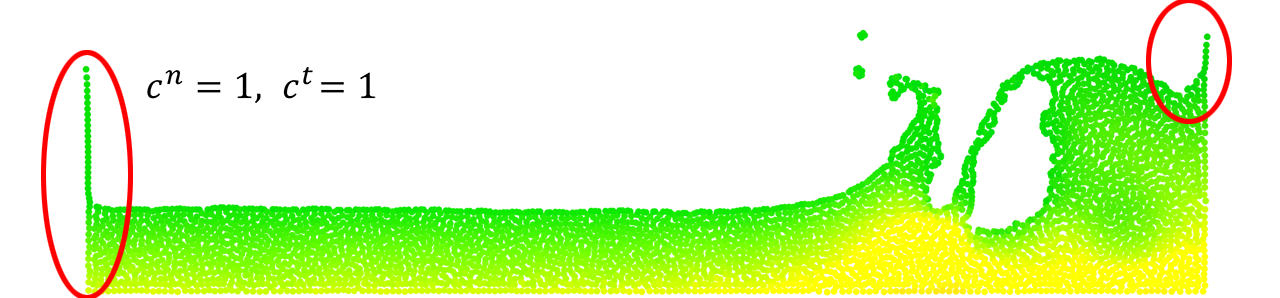}
	\caption{\label{fig:bTests} A two-dimensional dambreak test with four different solid wall boundary conditions. Significant differences of flow patterns can be noticed near the solid wall.}
\end{figure}
\textbf{Evaluation}
To evaluate the accuracy of our incompressible fluid solver, a benchmark test for hydrostatic water is performed.
A set of pressure calculations with three different initial heights are carried out as shown in Figure~\ref{fig:static}(a), where the water density is set to $1000kg/m^3$ and the gravity is set to $-9.8m/s^2$.
In the test, a no-slip solid wall boundary condition is imposed.
The particle shifting algorithm is temporarily neglected so we can evaluate the accuracy of the incompressibility solver alone.
Calculated pressure at the bottom of water shows that the simulation results converge to the analytical solution well, as shown in Figure~\ref{fig:static}(b).
Besides, it can also be noticed from the video that the total volumes for all three tests are well preserved.
In our second test, four different solid wall boundary conditions are imposed on a two-dimensional dambreak example.
We can notice significant differences in the pressure fields and flow patterns near the solid wall boundary, as shown in Figure~\ref{fig:bTests}.
Finally, Figure~\ref{fig:dambreak} compares the fluid patterns for different values of $\kappa$.
In the case of $\kappa = 0$, the particle shifting algorithm is simply not performed.
In other cases, it can be noticed that the particle shifting algorithm helps reduce pressure fluctuations caused by the irregularity of particle distributions.
Besides, stronger surface tension effects are captured with a larger value of $\kappa$.

\begin{figure}[t]
	\centering
	\subfigure[staggered SPH]{\includegraphics[width=0.495\linewidth]{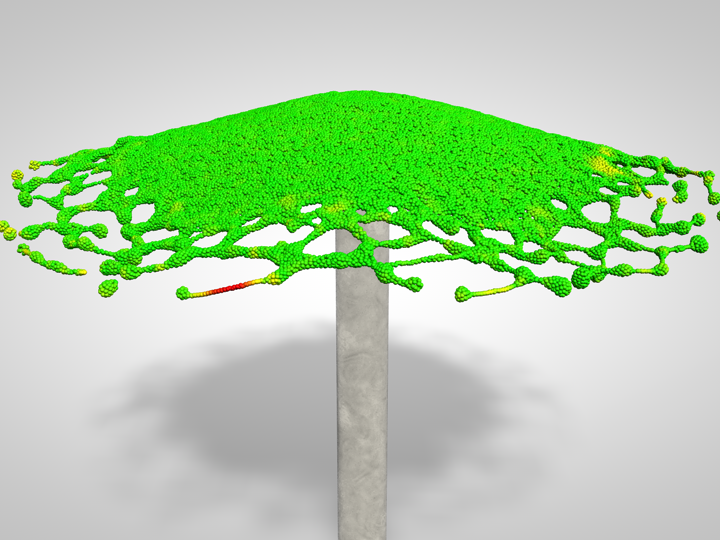}
	\includegraphics[width=0.495\linewidth]{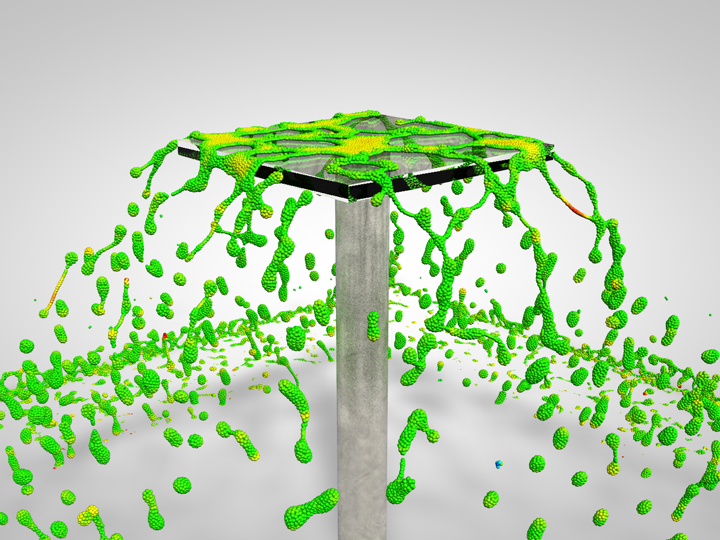}
	}
	\subfigure[our method]{\includegraphics[width=0.495\linewidth]{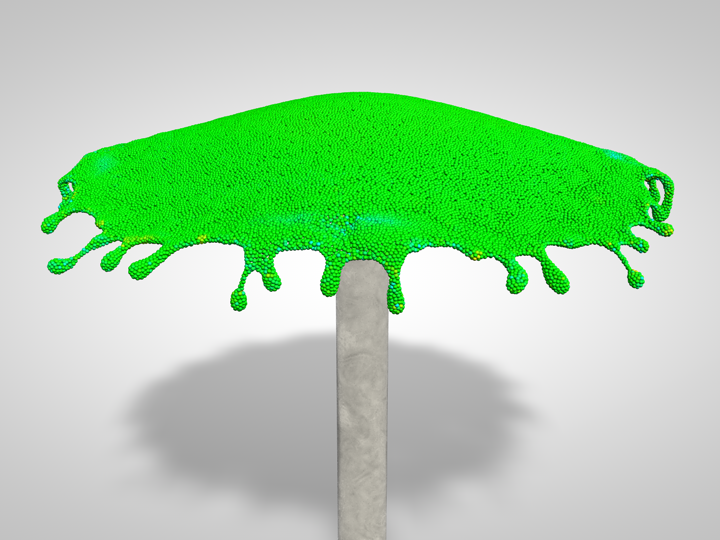}
	\includegraphics[width=0.495\linewidth]{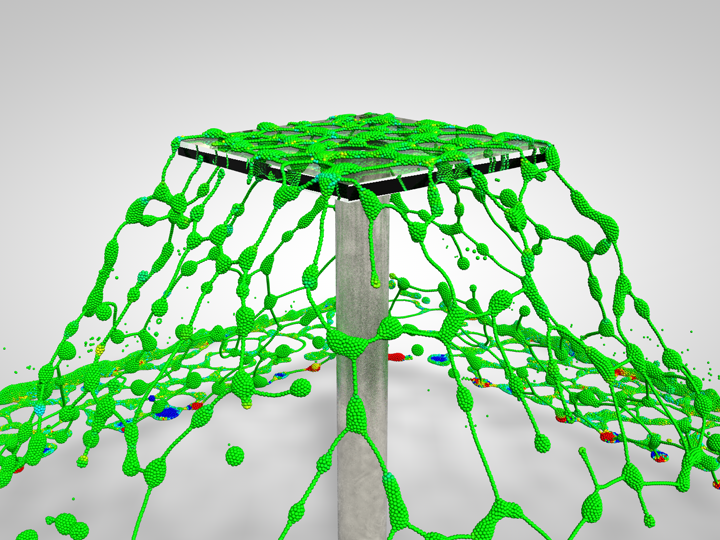}}
	\vspace{-0.1in}
	\caption{\label{fig:table}  A liquid ball onto a table. All other conditions for the comparison are the same except we use the position-based method~\cite{Macklin:2013:PBF}, which is not necessary with our method, to stabilize the simulation for \highlight{staggered SPH}. }
\end{figure}
\textbf{Comparison to staggered SPH~\cite{He:2012:SMS}}
Figure~\ref{fig:table} shows a comparison between our method and the staggered SPH method by dropping a liquid ball onto a table.
To perform a fair comparison, we apply our particle shifting algorithm for both methods.
Besides, we use the position-based method~\cite{Macklin:2013:PBF} to stabilize the simulation for \highlight{staggered SPH}, which is not necessary with our method.
It can be noted from the comparison that our method preserves thin features better than the \highlight{staggered SPH} method, especially for liquid jets and sheets.
One reason could be due to the sensitiveness of staggered SPH in imposing free boundary conditions on particles.
Another reason is that the Laplacian operator used in staggered SPH is quite similar to the one proposed by Cummins and Rudman~\cite{Cummins:1999:SPM}, therefore, the same problem exists as analyzed in Section 3.2.

\begin{figure}[t]
	\centering
	\subfigure[oblique view]{\includegraphics[height=2.5in]{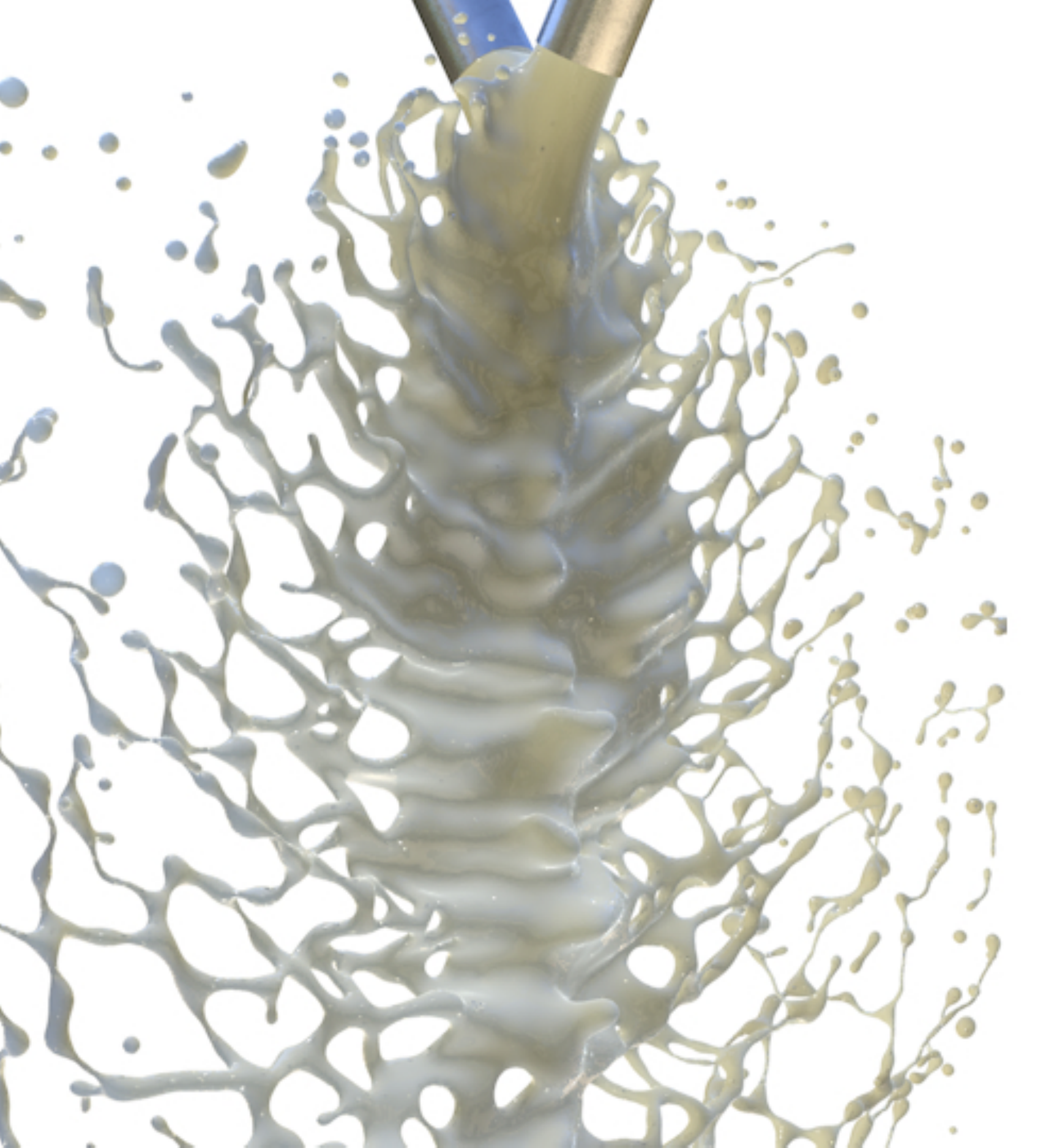}}
	\subfigure[side view]{\includegraphics[height=2.5in]{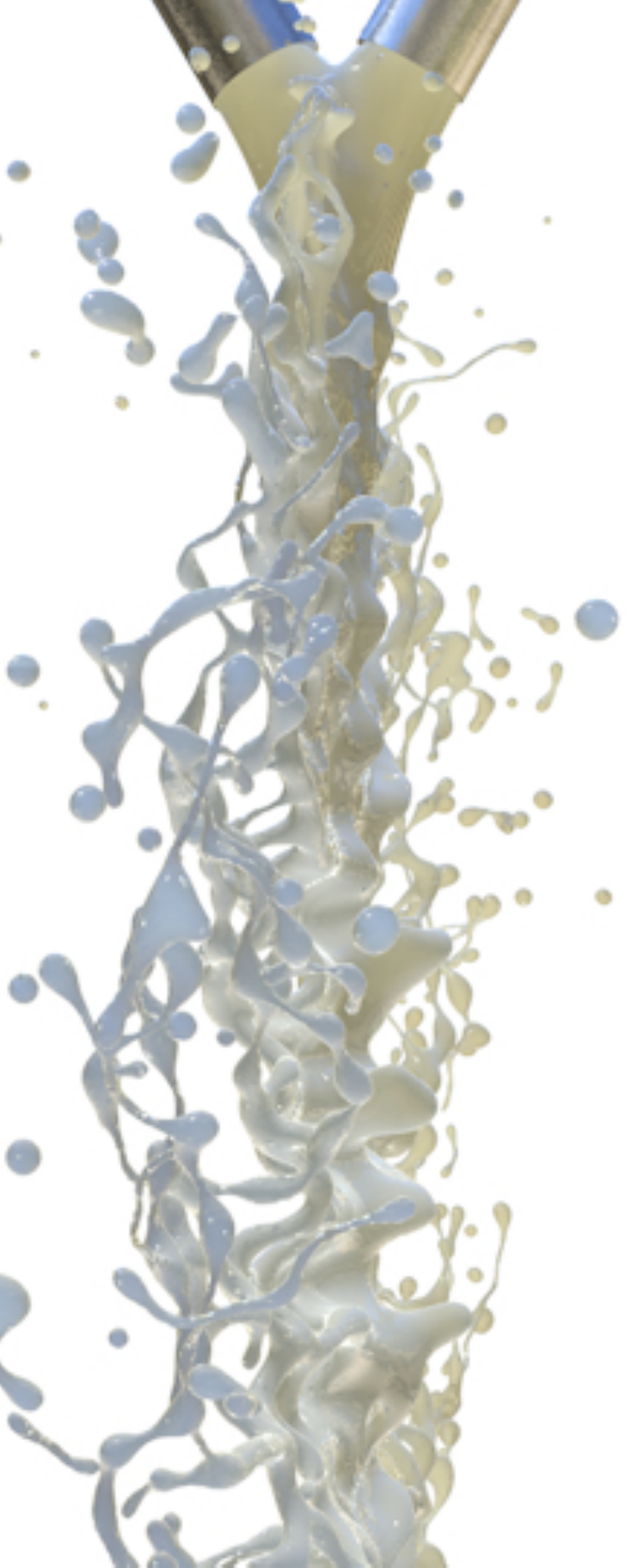}}
	\vspace{-0.1in}
	\caption{\label{fig:240} High-velocity impinging jets with an inlet velocity of $24m/s$. Impact waves arising from hydrodynamic instabilities are captured with our method.}
\end{figure}

\begin{figure*}[t]
	\centering
	\includegraphics[width=1.0\linewidth]{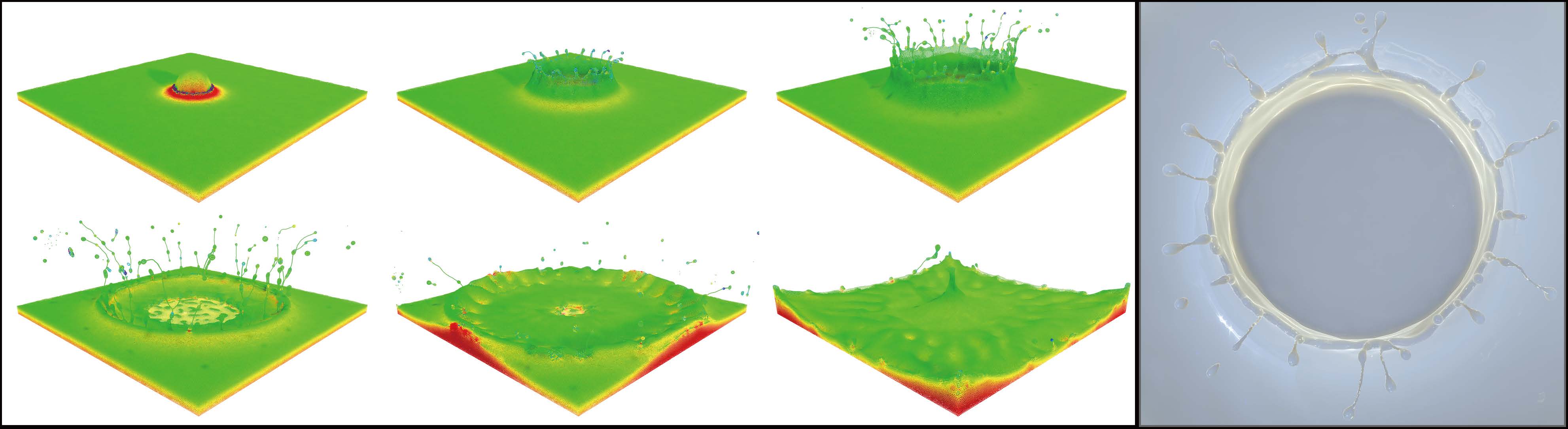}
	\vspace{-0.1in}
	\caption{\label{fig:milkcrown} Milk crown. This image shows the dynamic process of the formation and fragmentation of liquid jets and sheets, where particles are color-coded to demonstrate the pressure field.}
\end{figure*}
\textbf{Liquid jets.}
Figure~\ref{fig:fishbonecmp} shows two identical liquid jets impinging against each other, and a thin sheet resembling the shape of a fish bone forms around the intersection of the two jets.
The flow patterns formed by the two liquid jets is extremely challenging for previous particle methods to capture due to the particle deficiency and particle clumping problems.
It portrays seven typical snapshots by different methods.
\highlight{To be fair in comparison, we only replace the incompressibility solver for the top three examples compared to the simulation in Figure~\ref{fig:fishbonecmp}(g), while keeping all other parts (including particle shifting, surface tension and viscosity) unchanged.
Since both PCISPH and IISPH neglect the negative pressure, no thin liquid details can be captured.
This also explains the importance of negative pressure forces in generating richer liquid details.
DFSPH seems to be able to generate more details, however, no apparent viscous fingering structures can be noticed due to its low accuracy in solving the fluid incompressibility.
To check how the dynamic behavior is affected by surface tension model, the bottom row compares three different surface tension models with other parts being solved in the same way as the simulation in Figure~\ref{fig:fishbonecmp}(g).
To be fair, for each simulation, we tested more that five different surface tension coefficients and pick the best one in Figure~\ref{fig:fishbonecmp}(d), (e) and (f).
Unfortunately, no previous surface tension model is able to create high-fidelity simulations of the formation and fragmentation of liquid sheets formed by two impinging jets.
One reason could be because all three previous surface tension models are taken in an explicit manner, therefore the magnitude of surface tension force would be oscillatory and make the simulation results deteriorate.}
Finally, by taking a large inlet velocity, Figure~\ref{fig:240} shows the impact waves arising from hydrodynamic instabilities can also be captured by our method.
For more discussions on the impinging jets problem, we refer to the work by Chen et al.~\cite{Chen:2013:HFS}.

\begin{figure*}[t]
	\includegraphics[width=1.0\linewidth]{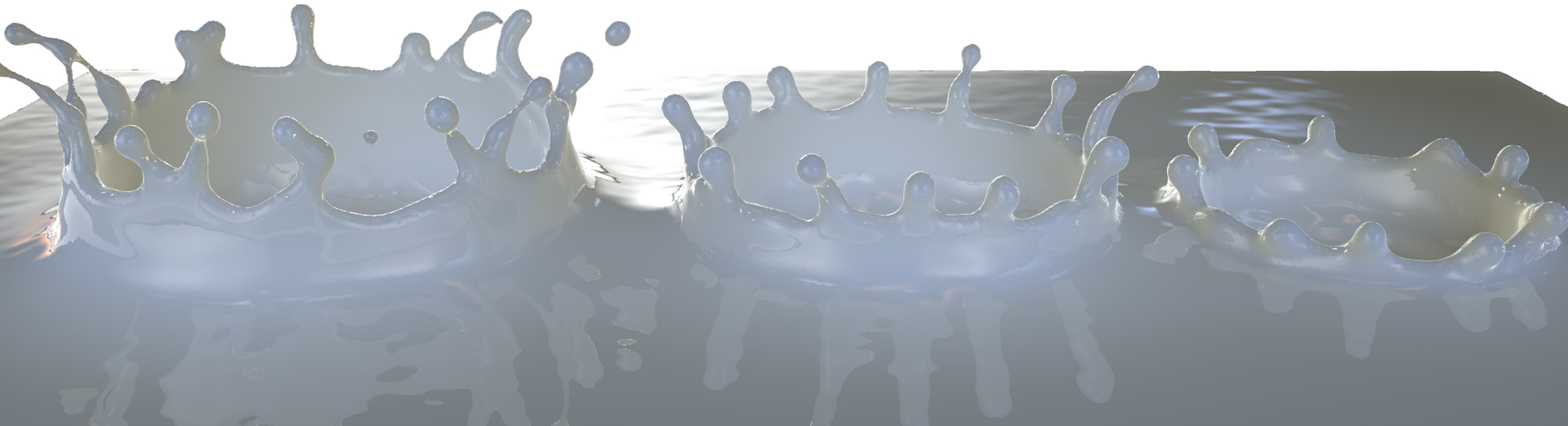}
	\caption{   \label{fig:teaser}
	Three milk droplets of different radii impacting on a milk body at $3m/s$. Under the combined action of pressure, surface tension and viscosity, thin jets are emitted around the rim that lead to the formation of milk crowns.
	}
\label{fig:three}
\end{figure*}

\textbf{Milk crown}. In Figure~\ref{fig:milkcrown}, we simulate a milk droplet that impacts on a milk body at $3 m/s$.
Under the action of pressure force and surface tension, several thin jets are emitted around the rim that subsequently lead to the formation of the milk crown.
Although the simulation result may still not comparable to those generated by the mesh-based methods~\cite{Thurey:2010:MAM}, we believe our method have made a great breakthrough over previous particle methods in modeling the fingering structures.
Figure~\ref{fig:three} additionally demonstrates three milk droplets with different radii that impacts on a milk body.
Different forms of the milk crown can be noticed due to the different size of milk droplets.

\begin{figure}[t]
	\centering
	\subfigure[Convergence examples of the conjugate gradient solver at different instants of time.]{\includegraphics[width=1.0\linewidth]{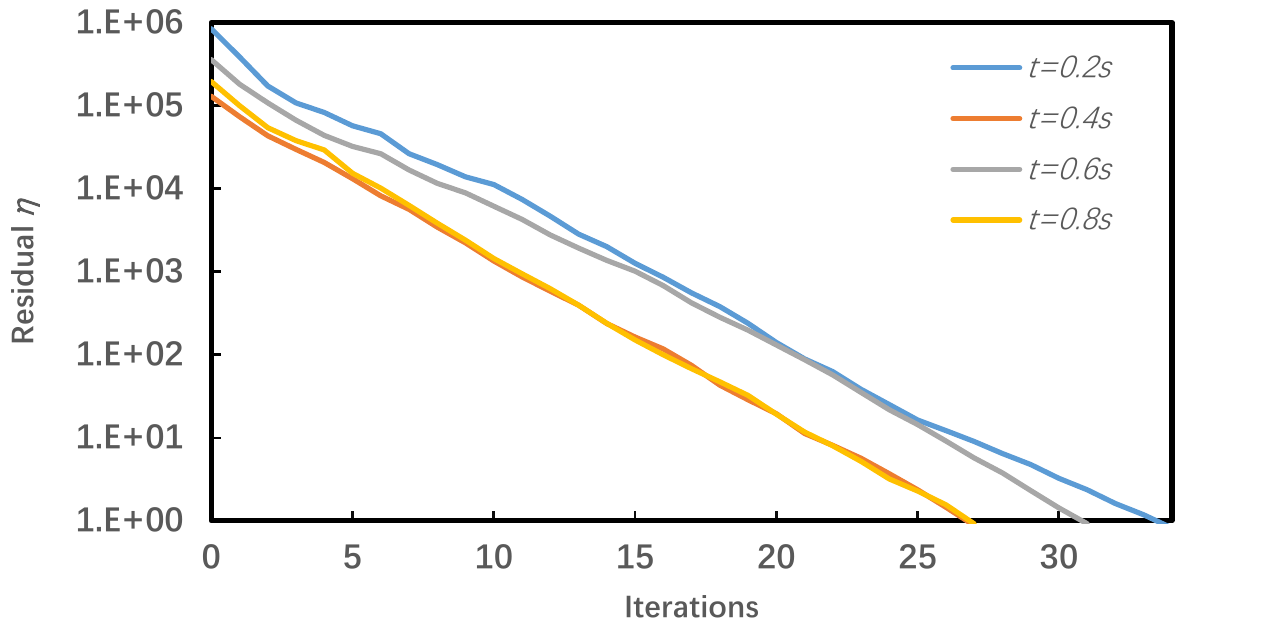}}
    \subfigure[Convergence examples of the particle shifting algorithm at different instants of time.]{\includegraphics[width=1.0\linewidth]{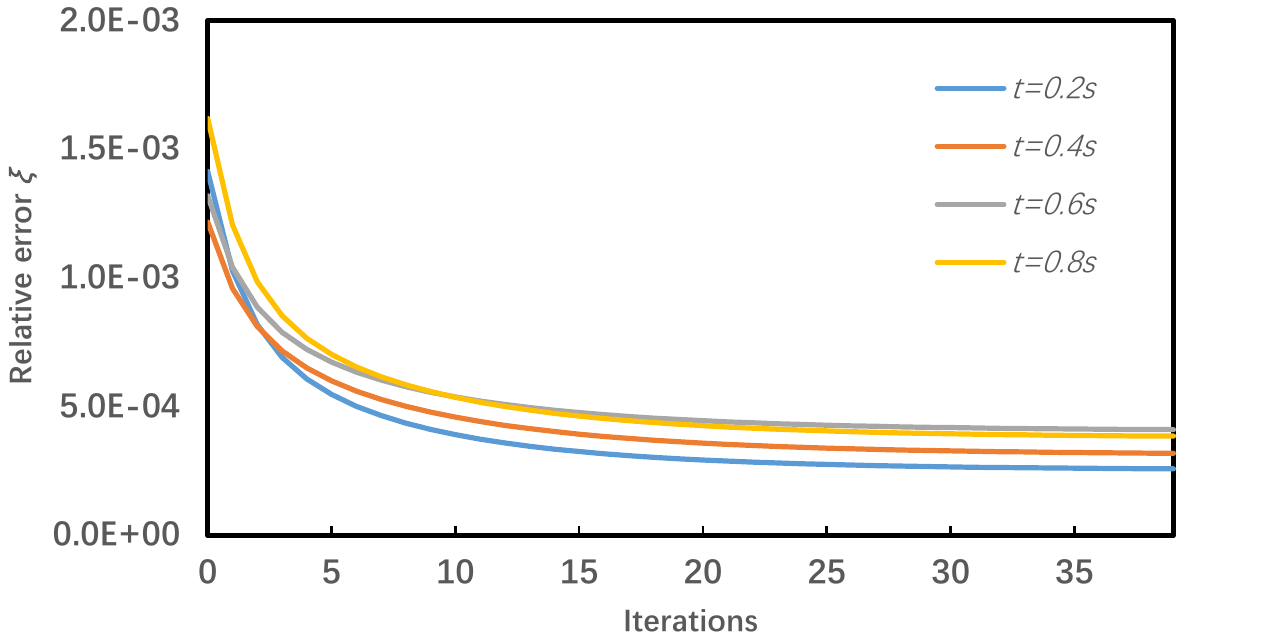}}
	\caption{\label{fig:convergence} Convergence statistics of the 2D dambreak example in Figure~\ref{fig:2ddambreak}(b). }
\vspace{-0.18in}
\end{figure}
\textbf{Performance}. The linear system of equations is solved with a conjugate gradient method.
By invoking the continuity equation for an ideal incompressible fluid, we have the following relationship that relates the residual of the linear system of equations to the density error
\begin{linenomath*}
	\begin{equation}
		\eta  = \frac{1}{N}\sum\limits_i {\left| {{{\mathcal L}_i} - {{\mathcal D}_i}} \right|} = \frac{1}{{\Delta t^2 N}}\sum\limits_i {\frac{{\left| {\Delta {\rho _i}} \right|}}{{{\rho _0}}}},
		\label{eq:error}
	\end{equation}
\end{linenomath*}
where $N$ is the total number of fluid particles.
For a 2D dambreak case, Figure~\ref{fig:convergence}(a) shows several examples of the convergence of the conjugate gradient solver.
It can be seen the residual error reduces by a factor of $10^6$ at around 30 iterations.
By requiring ${\left| {\Delta {\rho}} \right|} \le 10^{-3} {\rho _0}$, we can set the convergence condition as $\eta_0 = 10^{-3}/\Delta t^2$, therefore only around 15 iterations are required.

To measure the convergence rate for the particle shifting algorithm, we define a metric as follows
\begin{equation}
\xi  = \frac{1}{{{d_0}N}}\sum\limits_i {\left\| {\delta {\mathbf{x}_i}} \right\|}
\end{equation}
Figure~\ref{fig:convergence}(b) shows the convergence of the particle shifting algorithm.
It can be seen that the relative errors decrease rapidly at the first iterations.
Therefore, we typically take a constant number of 10 iterations for the particle shifting algorithm.
Table 1 shows other statistics and timings for all examples.

%\begin{figure}[t]
%\centering
%\includegraphics[width=1.0\linewidth]{images/sig.png}
%\vspace{-0.1in}
%  \caption{\label{fig:siggraph} The SIGGRAPH example.}
%\end{figure}

\textbf{Limitations}. Although the pressure force is corrected with $\beta_0$, the accuracy of our method is still less than a first-order accuracy, especially when the particle distribution is irregular.
Therefore, subtle fluctuations of the pressure field can be observed in solving the hydrostatic water test problem.
Besides, a slight momentum loss exists because the Taylor-series consistent pressure gradient model dose not fully conserve the momentum.
However, slightly losing a little momentum and kinetic energy is sometimes acceptable for computer graphics applications as long as it does not cause too much visual artifacts.
Finally, some efforts on parameter tuning is required to capture plausible fingering structures, among which the most important parameters are the inlet velocity, the gradient energy coefficient $\kappa$ and the viscosity coefficient.

\begin{table}[t]
\caption{Parameters and timings of all examples for one frame in average. }
	\centering
\vspace{-0.12in}
	{\begin{tabular}{l |c c c c c c c c}
			\toprule
			Name & size & $c^n$ & $c^t$ & $\kappa$ & time/frame \\
			\midrule
			Milk crown(Fig.~\ref{fig:milkcrown91}) & 1.2M & 0.2 & 0.0 & 0.6 & 26s \\
			3D dambreak(Fig.~\ref{fig:dambreak}) & 1.4M & 0.2 & 0.0 & 0$\sim$0.1 & 1.2 min \\
			Green vortex(Fig.~\ref{fig:vortex}) & 2209 & 1.0 & 1.0 & 0.1 & 0.26s \\
			2D dambreak(Fig.~\ref{fig:2ddambreak}) & 6589 & 0.2 & 0.0 & 0.1 & 0.48s \\
			Liquid on table(Fig.~\ref{fig:table}) & 113k & 0.0 & 0.0 & 0.6 & 3.5s \\
            Two jets(Fig.~\ref{fig:fishbonecmp}(g)) & 201k & - & - & 0.3 & 4.0s \\
            Three droplets(Fig.~\ref{fig:teaser}) & 1.9M & 0.2 & 0.0 & 0.6 & 1.7 min \\
			\bottomrule
	\end{tabular}}
	\label{tab:measurements}
\end{table}

\section{Conclusions and Future Work}
In this paper, we have introduced a novel approximate projection method under a variational staggered particle framework.
After setting up the discretized pressure Poisson equation and categorizing all fluid particles into four subsets, we solve the particle deficiency problem by analytically imposing free-surface boundary conditions for both the Laplacian operator and the source term.
Therefore, no ghost particles should be actually created during the simulation.
In calculating the pressure force, we address the particle clumping problem by extending a Taylor-series consistent pressure gradient model with kernel correction.
To regularize particle distributions, we introduce an iterative particle shifting algorithm motivated by Helmholtz free energy functional, which has the advantage of not only regularizing particle distributions, but also capturing plausible surface tension effects.

For our future work, we will first consider how to conserve the total momentum with the Taylor-series consistent pressure gradient model.
We will also investigate whether a higher order Laplacian~\cite{Khayyer:2012:HOL} or a higher order source~\cite{Khayyer:2009:MMP} can be integrated into our method to help improve the accuracy.
Besides, it would be interesting to investigate how to solve the particle clumping and particle deficiency problems for an adaptive SPH solver~\cite{Adams:2007:ASP,Solenthaler:2011:TPS,Winchenbach:2017:ICA}.
Finally, we will consider extending our method to handle more complex scenarios involving two-way coupling between fluid and solid.

\section{Acknowledgement}
\vspace{-0.06in}
The project was supported by the National Key R$\&$D Program of China (No.2017YFB1002700), the National Natural Science Foundation of China (No.6187070657, 61632003), Youth Innovation Promotion Association, CAS (No.2019109) and Key Research Program of Frontier Sciences, CAS (No. QYZDY-SSW-JSC041).

\vspace{-0.12in}
\ifCLASSOPTIONcaptionsoff
  \newpage
\fi



\bibliographystyle{IEEEtran}
\bibliography{VSISPH}



%\vfill


\end{document}